\documentclass[aps,pre,twocolumn,superscriptaddress,longbibliography,nofootinbib,showkeys,showpacs]{revtex4-2}

\usepackage[utf8]{inputenc}
\usepackage{amsfonts}
\usepackage{amsmath}
\usepackage{appendix}
\usepackage{bm}
\usepackage{braket}
\usepackage{caption}
\usepackage{color}
\usepackage{comment}
\usepackage{enumerate}
\usepackage{float}
\usepackage[top=2cm,left=2cm,right=2cm,bottom=2cm]{geometry}
\usepackage{graphicx}
\usepackage{hyperref}
\usepackage[utf8]{inputenc}
\usepackage{mathptmx} 
\usepackage{scalerel}
\usepackage{subcaption}
\usepackage{tikz}
\begin{document}
\title{Understanding the physics of D-Wave annealers: \\From Schrödinger to Lindblad to Markovian Dynamics}

\author{Vrinda Mehta\footnote{ORCID: 0000-0002-9123-7497}}
\affiliation{Institute for Advanced Simulation, J\"ulich Supercomputing Centre,\\
Forschungszentrum J\"ulich, D-52425 J\"ulich, Germany}
\author{Hans De Raedt\footnote{ORCID: 0000-0001-8461-4015}}
\affiliation{Institute for Advanced Simulation, J\"ulich Supercomputing Centre,\\
Forschungszentrum J\"ulich, D-52425 J\"ulich, Germany}

\author{Kristel Michielsen\footnote{ORCID: 0000-0003-1444-4262}}
\affiliation{Institute for Advanced Simulation, J\"ulich Supercomputing Centre,\\
Forschungszentrum J\"ulich, D-52425 J\"ulich, Germany}
\affiliation{RWTH Aachen University, 52056 Aachen, Germany}
\affiliation{AIDAS, 52425 J\"ulich, Germany}
\author{Fengping Jin\footnote{ORCID: 0000-0003-3476-524X}}
\affiliation{Institute for Advanced Simulation, J\"ulich Supercomputing Centre,\\
Forschungszentrum J\"ulich, D-52425 J\"ulich, Germany}
\begin{abstract}
    Understanding the physical nature of the D-Wave annealers remains a subject of active investigation. In this study, we analyze the sampling behavior of these systems and explore whether their results can be replicated using quantum and Markovian models. Employing the standard and the fast annealing protocols, we observe that the D-Wave annealers sample states with frequencies matching the Gibbs distribution for sufficiently long annealing times. Using Bloch equation simulations for single-qubit problems and Lindblad and Markovian master equations for two-qubit systems, we compare experimental data with theoretical predictions. Our results provide insights into the role of quantum mechanics in these devices.
\end{abstract}

\date{\today}
\maketitle

\section{Introduction}

How crucial is quantum theory in describing the results of D-Wave annealers? Put differently, can the data produced by these systems be explained or replicated using non-quantum models? If so, to what extent can these machines truly be considered ``quantum"? 
These questions have been lying around since the introduction of the D-Wave annealers~\cite{Harris2010,johnson2010scalable,johnson2011quantum,karimi2012investigating,boixo2014evidence,shin2014quantum,ronnow2014defining,lanting2014entanglement, albash2015reexamination, denchev2016computational,ronnow2014defining,maditaflux2020,king2022,king2023quantum,Kingbeyond2025,Vodeb25,albash2018adiabatic}.

The role of quantum phenomena in determining the performance of D-Wave quantum annealers has been a subject of intense debate since their introduction \cite{johnson2011quantum, boixo2014evidence, shin2014quantum}. A fundamental question is whether the observed behavior of these devices can be explained using non-quantum models or whether quantum effects, such as entanglement and tunneling, have to be incorporated to reproduce experimental results. Addressing this question is crucial for understanding the extent to which these machines leverage quantum mechanics for computational advantage \cite{ronnow2014defining, denchev2016computational, albash2018demonstration,Kingbeyond2025}.

Given the complexity of these large-scale annealers, an effective approach to investigate their behavior is to focus on minimal problems that retain the essential characteristics of these systems while being amenable to analytical and numerical methods. As demonstrated in our previous work~\cite{reverseannealing}, in theory, even 1- and 2-qubit problems can capture key features of the underlying dynamics, including the role of quantum coherence and thermal effects.  Indeed, these small problem instances have been an excellent testbed for evaluating the quantum nature of these devices \cite{lanting2014entanglement, albash2015reexamination}.

In this work, we analyze the statistics of the states sampled by D-Wave quantum annealers for specific problem instances. We study the frequency distributions of these sampled states, construct theoretical models, investigate the dynamical evolution of these systems by employing numerical simulations based on the Schr{\"o}dinger equation, the Lindblad master equation~\cite{Lindblad1976}, and the Markovian master equation, and scrutinize the extent to which the experimental data aligns with the different simulation results.

In the same vein as with the reverse annealing protocol in our previous work~\cite{reverseannealing}, the present work focuses on understanding the behavior of the D-Wave quantum annealers using the standard and the recently introduced fast annealing protocols.

Our results contribute to the broader discussion on the quantum nature of D-Wave devices and the extent to which their behavior can be emulated by classical processes. By systematically comparing theoretical predictions with experimental data, we provide insights into the role of quantum effects in quantum annealing and their implications for practical applications.

It is important to emphasize that the present work explores regimes that are complementary to those studied in Refs.~\cite{king2022,king2023quantum,Kingbeyond2025}, particularly in terms of problem complexity and scale. First, our study focuses on a fundamental aspect—the frequencies with which the D-Wave annealer samples different states—and compares them to theoretical expectations from Schr{\"o}dinger, Lindblad, and Markovian dynamics. This approach offers a raw and direct perspective on the system’s behavior. Second, our analysis is centered on 1- and 2-spin systems, which allow for a more controlled and detailed examination of the underlying physical processes. These small-scale problems provide the best opportunity to observe signatures of quantum coherence; failure to detect such effects at this level would significantly diminish the prospects of observing them in larger, more complex systems.

The paper is structured as follows: Section~\ref{sec:methods} outlines the methods and problems considered in this study. Section~\ref{sec:motivation} presents the results obtained from the D-Wave annealer, while Section~\ref{sec:simulation} discusses the corresponding simulation results. Finally, Section~\ref{sec:conclusion} summarizes the key findings of the paper.

\section{Methods and problems}   
\label{sec:methods}
\label{sec:probs}
 
The Hamiltonian for the quantum annealing process, implemented on the D-Wave annealers, is given by \cite{Dwave}
\begin{align}
    \frac{H(t)}{\hbar} &= \frac{\pi A(s=t/t_a)}{h} H_D + \frac{\pi B(s=t/t_a)}{h} H_P, \nonumber\\
    H_D &=-\sum_i\sigma_i^x, \nonumber\\
    H_P &= \sum_i h_i\sigma_i^z + \sum_{i>j} J_{ij}\sigma_i^z\sigma_j^z,
    \label{eq:DWHamil}
\end{align}
where $t_a$ is the annealing time, $\sigma_i^x$, $\sigma_i^z$ are the Pauli matrices, $h_i$ are effective local fields, and $J_{ij}$ are the effective coupling terms between qubits $i$ and $j$.
Functions $A(s)$ and $B(s)$ (expressed in units GHz) define the annealing schedules and are vastly different for the standard and the fast annealing protocols (see appendix~\ref{app:sched}). Under the conditions that $A(0)\gg B(0)$, $A(1)\ll B(1)$, and that $t_a$ is sufficiently large, the adiabatic theorem \cite{born1928beweis,kato1950adiabatic} guarantees that the system will find the ground state of the problem Hamiltonian $H_P$, if the system is initialized in the ground state
\begin{equation}
    \ket{++ \cdots +}_N = \frac{1}{\sqrt{2}^N}(\ket{0}+\ket{1})^{\otimes N}\;,
    \label{eq:initialstate}
\end{equation}
of the driver Hamiltonian $H_D$, where $N$ is the number of qubits.

The fast annealing protocol requires that the fields in Eq.~(\ref{eq:DWHamil}) have to be zero \cite{Dwave}. A workaround for implementing problems with non-zero $h_i$'s is to use an additional qubit for every variable that has a non-zero $h_i$~\cite{Dwave}. The state of the additional qubit is set by applying a fixed flux bias offset to it. Then, the coupling $J_{ij}=h_i$ between the original qubit $i$ and the additional qubit $j$ effectively implements the original problem. For a 1-spin problem with a magnetic field $h_1^z$, the resulting annealing Hamiltonian reads
\begin{align}
    \frac{H'(t)}{\hbar} = - \frac{\pi A(s)}{h}~ (\sigma_1^x+\sigma_2^x)+ \frac{\pi B(s)}{h}~J_{12}\sigma_1^z \sigma_2^z + h_{FB}\sigma_2^z\;,
\end{align}
where $J_{12}=h_1^z$ and $h_{FB} \gg \pi\max(A(s),B(s)h_1)/h$ is the constant flux bias offset applied to the second qubit. 

Similarly, for a given 2-spin problem with fields $h_1$ and $h_2$, and coupling $J$, the corresponding annealing Hamiltonian for the fast annealing protocol reads \begin{align}
\frac{H''(t)}{\hbar} =& -\frac{\pi A(s)}{h}~ (\sigma_1^x+\sigma_2^x+\sigma_3^x+\sigma_4^x)+ h_{FB}(\sigma_3^z+\sigma_4^z)\nonumber\\
&+ \frac{\pi B(s)}{h}~(J_{12}\sigma_1^z \sigma_2^z+J_{13}\sigma_1^z \sigma_3^z+J_{24}\sigma_2^z \sigma_4^z) \;,
\end{align}
where $J_{13}=h_1$ and $J_{24}=h_2$.

We study the annealing time dependence of the probabilities of all the relevant energy levels, assigned by using their sampling frequencies. All the results presented in this paper are obtained by averaging over multiple runs across different qubits available in the D-Wave systems. Additionally, these multiple runs are performed either sequentially—where only one instance of the problem is executed at a time—or simultaneously—where multiple instances are submitted in parallel.

To numerically study the annealing process of the D-Wave systems, we use the Bloch equations, the Lindblad master equation, and the Markovian master equation. We use the functions obtained by fitting to the D-Wave annealing schedule data for the standard and, in addition, the newly introduced fast annealing protocols (see appendix~\ref{app:sched}).

We study the same simple yet diverse problem set as the one in Ref.~\cite{reverseannealing}. Specifically, we examine the 1- and 2-spin problem instances described in Ref.~\cite{reverseannealing}, as well as ferromagnetic spin chains. The 2-variable problem instances used in this paper are as follows.
\begin{itemize}
    \item 2S1: $h_1=-1.00$, $h_2=-1.00$, $J_{12}=0.95$
    \item 2S2: $h_1=-1.00$, $h_2=-1.00$, $J_{12}=-1.00$
    \item 2S3: $h_1=-0.95$, $h_2=-0.95$, $J_{12}=1.00$
\end{itemize}

\section{D-Wave results}
In this section, we show representative results obtained by executing the standard and the fast annealing protocols on the D-Wave annealers.

While we mainly focus on the results from the D-Wave Advantage\_5.4 system in this paper, the results from the Advantage\_4.1 system show similar behavior. The D-Wave data reported here is reproducible and systematic.
\label{sec:motivation}
\begin{figure*}
     \begin{minipage}{0.49\textwidth}
         \centering
         \includegraphics[scale=0.6]{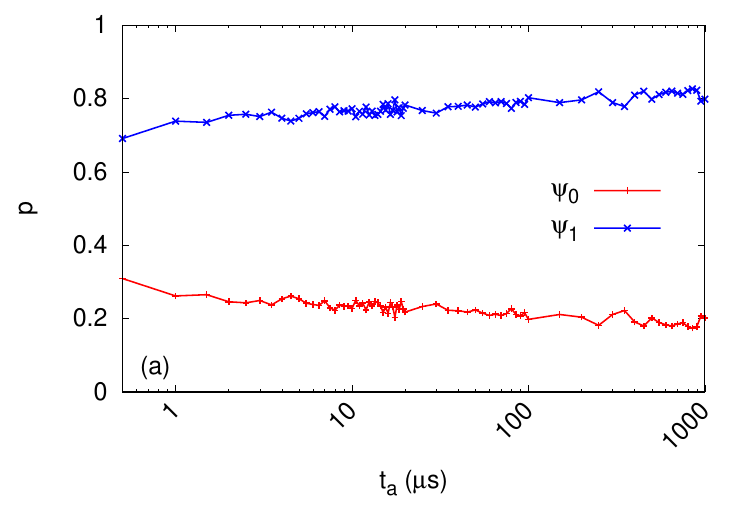}
         \put(-70, 93){\colorbox{white}{\makebox(21,6){\footnotesize $\ket{\uparrow}$}}}
    \put(-70, 83){\colorbox{white}{\makebox(21,6){\footnotesize $\ket{\downarrow}$}}}
     \end{minipage}
     \hfill
     \begin{minipage}{0.49\textwidth}
         \centering
         \includegraphics[scale=0.6]{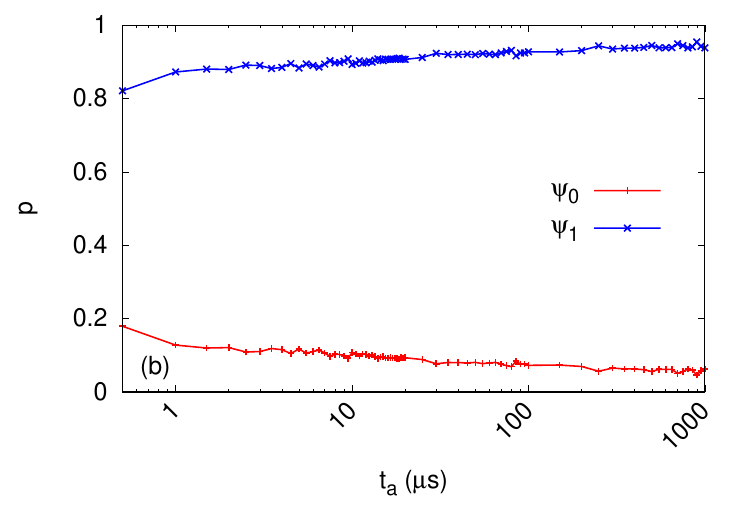}
         \put(-70, 93){\colorbox{white}{\makebox(21,6){\footnotesize $\ket{\uparrow}$}}}
    \put(-70, 83){\colorbox{white}{\makebox(21,6){\footnotesize $\ket{\downarrow}$}}}
     \end{minipage} \\
     \hfill
     \caption{(Color online) D-Wave data obtained using the standard annealing protocol for a 1-spin problem with 
     (a) $h_1=0.1$ and (b) $h_1=0.2$.}
    \label{fig:motiv_1spin}
\end{figure*}

\begin{figure*}
     \begin{minipage}{0.33\textwidth}
         \centering
         \includegraphics[width=\textwidth]{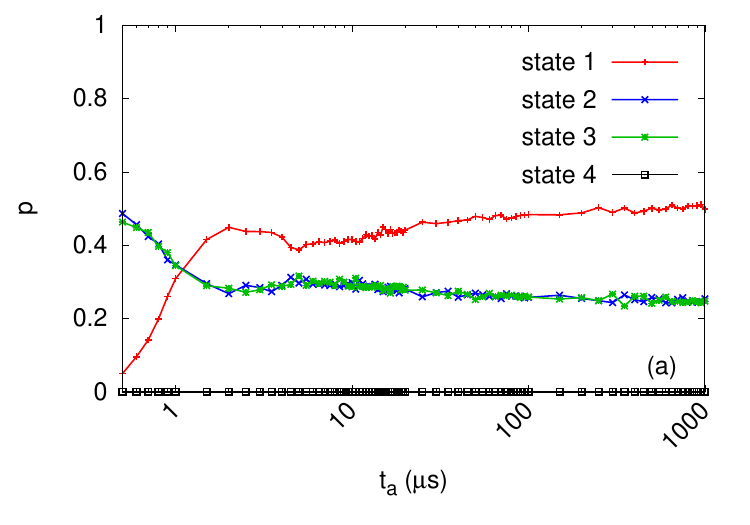}
         \put(-51, 100){\colorbox{white}{\makebox(12,2){\footnotesize $\ket{\uparrow\uparrow}$}}}
         \put(-51, 92){\colorbox{white}{\makebox(12,2){\footnotesize $\ket{\uparrow\downarrow}$}}}
         \put(-51, 84){\colorbox{white}{\makebox(12,2){\footnotesize $\ket{\downarrow\uparrow}$}}} 
         \put(-51, 75){\colorbox{white}{\makebox(12,2){\footnotesize $\ket{\downarrow\downarrow}$}}}
     \end{minipage}
     \hfill
     \begin{minipage}{0.33\textwidth}
         \centering
         \includegraphics[width=\textwidth]{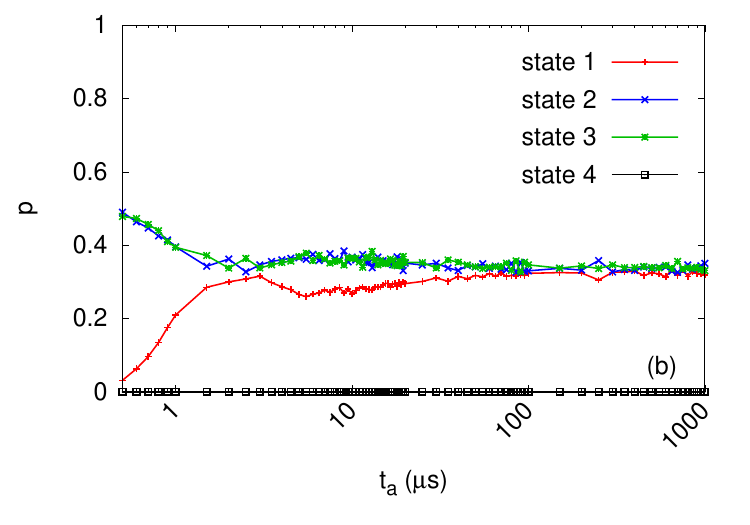}
         \put(-51, 100){\colorbox{white}{\makebox(12,2){\footnotesize $\ket{\uparrow\uparrow}$}}}
         \put(-51, 92){\colorbox{white}{\makebox(12,2){\footnotesize $\ket{\uparrow\downarrow}$}}}
         \put(-51, 84){\colorbox{white}{\makebox(12,2){\footnotesize $\ket{\downarrow\uparrow}$}}} 
         \put(-51, 75){\colorbox{white}{\makebox(12,2){\footnotesize $\ket{\downarrow\downarrow}$}}}
     \end{minipage}
        \begin{minipage}{0.33\textwidth}
         \centering
         \includegraphics[width=\textwidth]{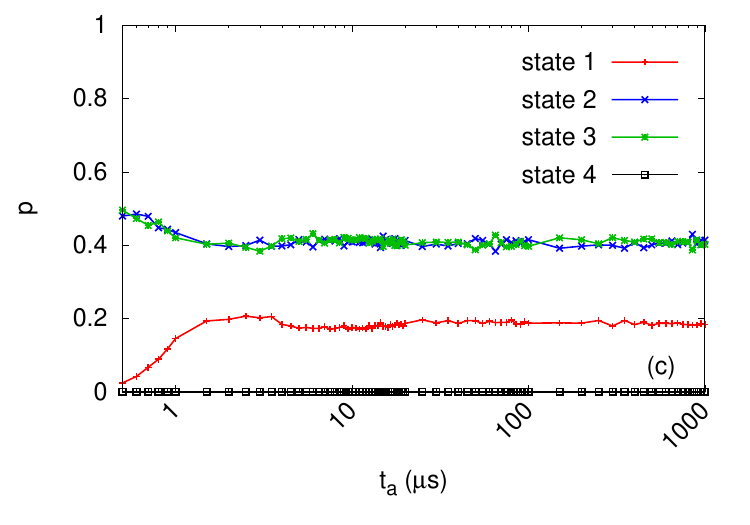}
         \put(-51, 100){\colorbox{white}{\makebox(12,2){\footnotesize $\ket{\uparrow\uparrow}$}}}
         \put(-51, 92){\colorbox{white}{\makebox(12,2){\footnotesize $\ket{\uparrow\downarrow}$}}}
         \put(-51, 84){\colorbox{white}{\makebox(12,2){\footnotesize $\ket{\downarrow\uparrow}$}}} 
         \put(-51, 75){\colorbox{white}{\makebox(12,2){\footnotesize $\ket{\downarrow\downarrow}$}}}
     \end{minipage}
     \hfill
     \caption{(Color online) D-Wave data obtained using the standard annealing protocol for a 2-spin problem instances with 
     (a) 2S1, (b) 2S2, and (c) 2S3.}
    \label{fig:motiv_2spin}
\end{figure*}
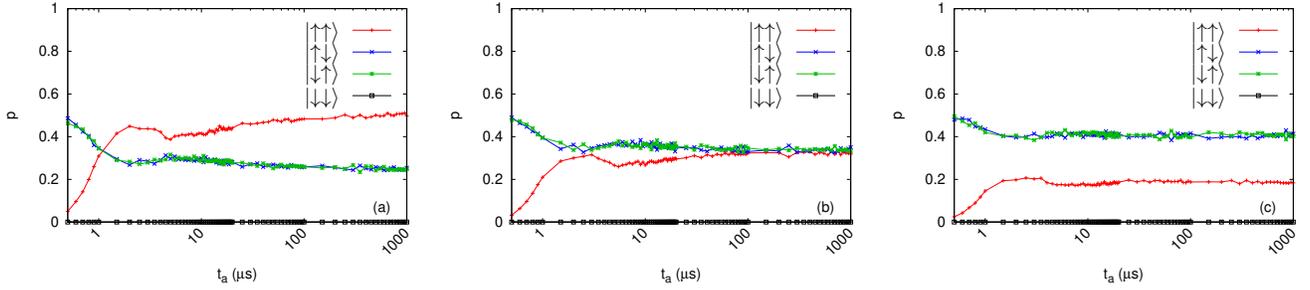
\subsection{Standard quantum annealing}
Figure~\ref{fig:motiv_1spin} shows the results for the 1-spin problem with $h_1=0.1$ and $h_1=0.2$ for different values of $t_a$. We note that with an increasing annealing time $t_a$, the probabilities $p(t_a)$ tend to stabilize. For the case with $h_1=0.1$, $p_{\uparrow}\approx0.2$ and $p_{\downarrow}\approx0.8$ for annealing time $t_a=1000~\mu$s on the quantum annealer, while for $h_1=0.2$, $p_{\uparrow}\approx0.06$ and $p_{\downarrow}\approx0.94$. The corresponding $p(t_a=1000~\mu \mathrm{s})$ for the states $\ket{\uparrow}$ and $\ket{\downarrow}$ for the 1-spin problem with $h_1=0$ (results not shown) are close to $1/2$ for different values of $t_a$.

\begin{figure*}
     \begin{minipage}{0.5\textwidth}
         \centering
         \includegraphics[scale=0.6]{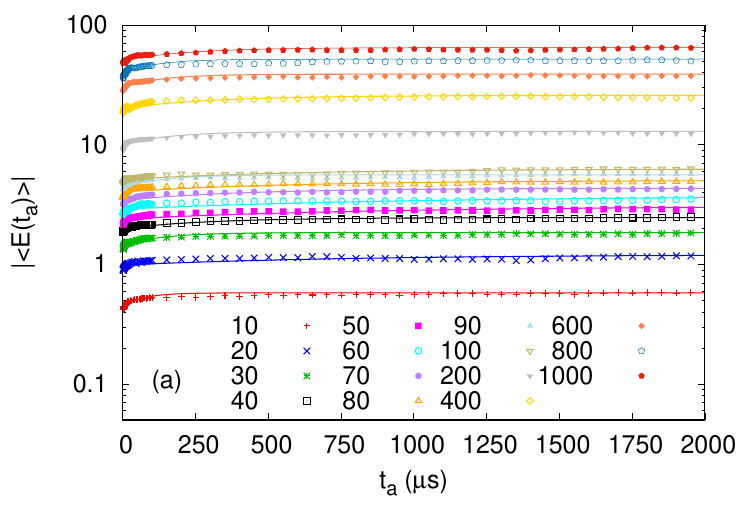}
     \end{minipage}
     \hfill
     \begin{minipage}{0.49\textwidth}
         \centering
         \includegraphics[scale=0.6]{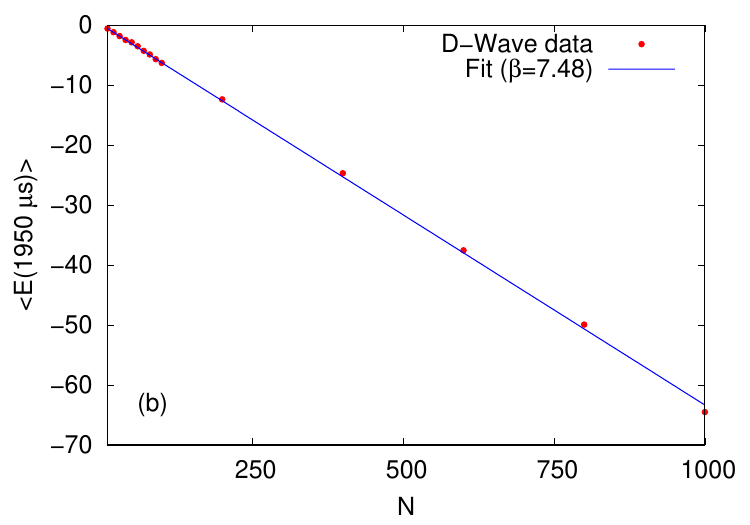}
     \end{minipage} \\
     \hfill
     \caption{(Color online) D-Wave data for the ferromagnetic spin chain of length $N=10,\ldots,1000$. (a) Absolute value of the average energy $\langle E(t_a)\rangle$ as a function of annealing time $t_a$ and (b) the average energy at $t_a=1950~\mu$s as a function of chain length $N$.}
    \label{fig:ferrochain_std}
\end{figure*}

Next, in Fig.~\ref{fig:motiv_2spin}, we show the D-Wave data for the three 2-spin instances. Once again, we find the probabilities $p(t_a)$ tending to a stable value for long annealing times, even if not as stabilized as by using the reverse annealing protocol \cite{reverseannealing}. 

A persistent, unexpected feature that accompanies these scans as a function of $t_a$ is that the value of $p(t_a)$ for $t_a \leq 2~\mu$s, where we find that the probability for state $\ket{\uparrow\uparrow}$ starts from a value close to zero till it eventually reaches relatively close to the stable value. Consequently, the probabilities for the states $\ket{\uparrow\downarrow}$ and $\ket{\downarrow\uparrow}$ are larger for these annealing times than their stable values for the larger annealing times. We will refer to this feature as the ``dips-and-bump", respectively, in the subsequent parts of this paper. Since even the annealing time $t_a=0.5~\mu$s is more than sufficient for the success probability of ideal quantum annealing to be close to one for problems of this size, the success probability being close to zero demands an explanation.

Given the results for these simple problems and our conclusions for reverse annealing \cite{reverseannealing}, we expect that the probabilities $p(t_a)$ eventually tend to their equilibrium values given by 
\begin{equation}
    p_i = \frac{g_i e^{-\beta E_i}}{\sum_ig_i e^{-\beta E_i}},
    \label{eq:equil}
\end{equation}
where $g_i$ and $E_i$ are the degeneracy and the energy, respectively, of the $i$th level of the problem Hamiltonian, with $E_{i+1} > E_i$, and $\beta = C/T$ for $C = h\, B(s=1)/(2 k_B)\times{10}^{9}=0.206~\mathrm{K}$ and some effective temperature $T$ (expressed in kelvin). 
The computational basis states $\ket{\uparrow}$ and $\ket{\downarrow}$ correspond to the spin eigenvalues
$S=+1$ and $S=-1$, respectively, and are related to the index $i$ through $i=(1-S_1)/2$ for a 1-spin and $i=(1-S_1)/2+(1-S_2)$
for the 2-spin systems.

For testing the equilibration idea, we determine the value of $\beta$ for the 1-spin problem with $h_1=0.1$ and use the obtained value to calculate the corresponding values of $p_i$ for the other problems. From Fig.~\ref{fig:motiv_1spin}(a), we find $\beta=6.93$, which corresponds to an effective temperature of $T=29.7$~mK, of similar order of magnitude as the cryogenic temperature 13~mK of the D-Wave Advantage\_5.4 system. Moreover, this value of $\beta$ is the same as that obtained from the waiting time scan of this problem using reverse annealing \cite{reverseannealing}.
For the 1-spin problem with $h_1=0.2$, $\beta=6.93$, Eq.~(\ref{eq:equil}) yields $p_{\uparrow}=0.06$ and $p_{\downarrow}=0.94$, values that closely match the  frequencies obtained from the annealer. 

Considering the three 2-spin instances next, we find the same equilibrium probabilities as in reverse annealing~\cite{reverseannealing}. More specifically, for 2S1, this results in $p_{\uparrow\uparrow}= 0.50$, $p_{\uparrow\downarrow}=p_{\downarrow\uparrow}=0.25$, and $p_{\downarrow\downarrow}=0$. For 2S2, we obtain $p_{\uparrow\uparrow}=p_{\uparrow\downarrow}=p_{\downarrow\uparrow}=0.33$ and $p_{\downarrow\downarrow}=0$, while for 2S3, Eq.~(\ref{eq:equil}) yields $p_{\uparrow\uparrow}= 0.20$,  $p_{\uparrow\downarrow}=p_{\downarrow\uparrow}=0.40$, and $p_{\downarrow\downarrow}=0$. The corresponding D-Wave data agree very well with these theoretical values, suggesting that the D-Wave annealers tend towards thermal equilibrium for sufficiently long annealing times. 

To consolidate these conclusions, we move to the ferromagnetic spin chains with up to $N=1000$ variables and the homogeneous coupling $J=-0.1$. In Fig.~\ref{fig:ferrochain_std}(a), we show the mean energy values of the samples obtained from the annealer as a function of annealing times $t_a$ for various problem sizes. Panel (b) of the figure shows the mean energy $\langle E(1950~\mu s) \rangle$ as a function of the problem size. As the energies and the degeneracies of all the energy levels of this problem are known analytically, we fit 
\begin{equation}
    \langle E \rangle = \frac{\sum_i g_i E_i e^{-\beta E_i}}{\sum_ig_i e^{-\beta E_i}} = -J(N-1) \tanh \beta J,
    \label{eq:equil_energy}
\end{equation}
to the empirical data via the parameter $\beta$. The excellent agreement of the resulting fit with the data, along with a value of $\beta = 7.48$, which is close to the one found earlier, corroborates the conjecture of relaxation to the equilibrium  distribution. A similar procedure, when applied using the fast annealing protocol (results not shown), yields nearly identical results.

Recall that the results presented above are obtained by averaging over multiple sequential runs.
While similar behavior is observed at the individual run level, we observe clear systemic biases in the data (not shown in the paper). A careful consideration is therefore required when interpreting their output, as discussed in Appendix~\ref{app:maxent}. However, choosing the simultaneous mode of collecting data is found to introduce an overall bias in the results, which can be mitigated using spin-reversal transformations, as detailed in Appendix~\ref{app:simultaneous}.

\subsection{Fast annealing}

The results presented in the previous section, obtained using standard quantum annealing, show no recognizable signs of coherent evolution. This begs the question: What happens for shorter annealing times than those permitted for the standard annealing protocol?
We address this question using the recently introduced fast annealing feature~\cite{Dwave}, focusing on the 1- and 2-spin instances. 

\begin{figure}[H]
    \centering
    \includegraphics[scale=0.6]{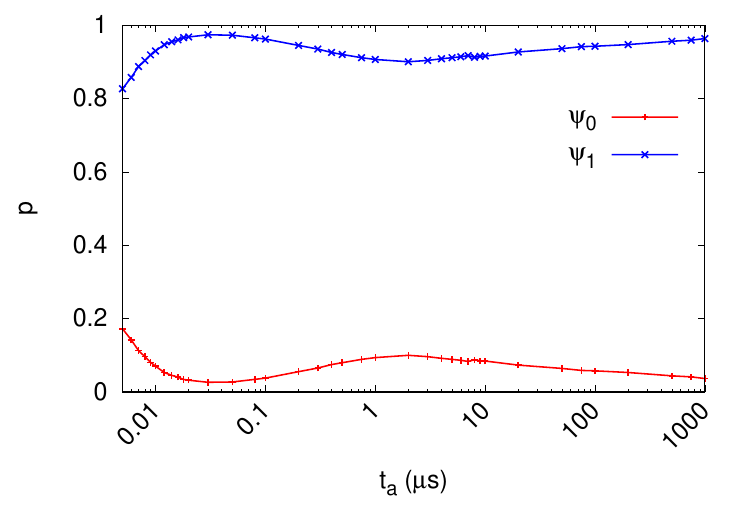}
    \put(-67, 114){\colorbox{white}{\makebox(21,6){\footnotesize $\ket{\uparrow}$}}}
    \put(-67, 104){\colorbox{white}{\makebox(21,6){\footnotesize $\ket{\downarrow}$}}}
    \caption{(Color online) D-Wave data obtained using the fast annealing protocol for 1-spin problem with  $h_1=0.25$.}
    \label{fig:fast1spin}
\end{figure}

The resulting data for the 1-spin problem with $h=0.25$ is shown in Fig.~\ref{fig:fast1spin}. In contrast to the data obtained using standard quantum annealing, in this case, we find that for the initial times ($t_a \lesssim 20$~ns), the ground state probability increases with increasing the annealing time. This can be a signature of an underlying coherent evolution of the initial state, which necessitates further investigation (see below). On further increasing the annealing time, the probabilities tend to relax to their stationary values.

\begin{figure*}
     \begin{minipage}{0.33\textwidth}
         \centering
         \includegraphics[width=\textwidth]{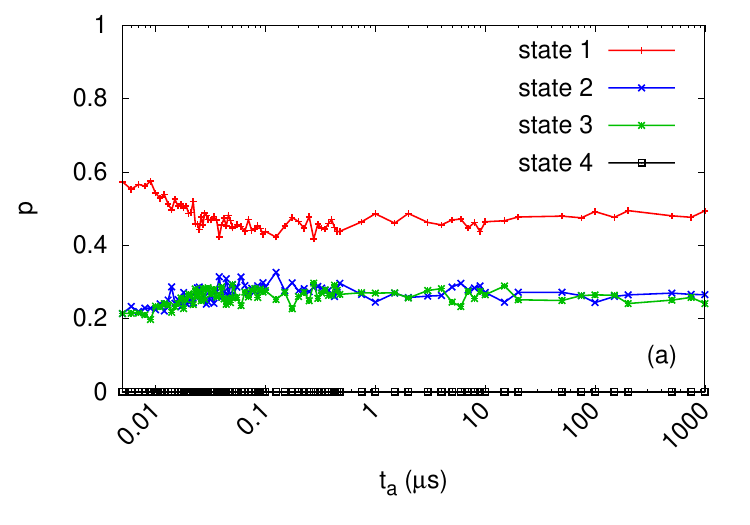}
         \put(-51, 102){\colorbox{white}{\makebox(12,2){\footnotesize $\ket{\uparrow\uparrow}$}}}
         \put(-51, 94){\colorbox{white}{\makebox(12,2){\footnotesize $\ket{\uparrow\downarrow}$}}}
         \put(-51, 86){\colorbox{white}{\makebox(12,2){\footnotesize $\ket{\downarrow\uparrow}$}}} 
         \put(-51, 78){\colorbox{white}{\makebox(12,2){\footnotesize $\ket{\downarrow\downarrow}$}}}
     \end{minipage}
     \hfill
     \begin{minipage}{0.33\textwidth}
         \centering
         \includegraphics[width=\textwidth]{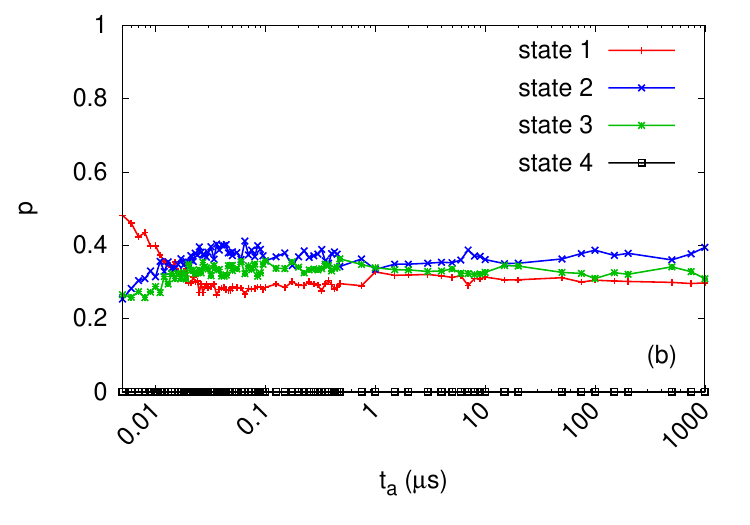}
         \put(-51, 102){\colorbox{white}{\makebox(12,2){\footnotesize $\ket{\uparrow\uparrow}$}}}
         \put(-51, 94){\colorbox{white}{\makebox(12,2){\footnotesize $\ket{\uparrow\downarrow}$}}}
         \put(-51, 86){\colorbox{white}{\makebox(12,2){\footnotesize $\ket{\downarrow\uparrow}$}}} 
         \put(-51, 78){\colorbox{white}{\makebox(12,2){\footnotesize $\ket{\downarrow\downarrow}$}}}
     \end{minipage}
        \begin{minipage}{0.33\textwidth}
         \centering
         \includegraphics[width=\textwidth]{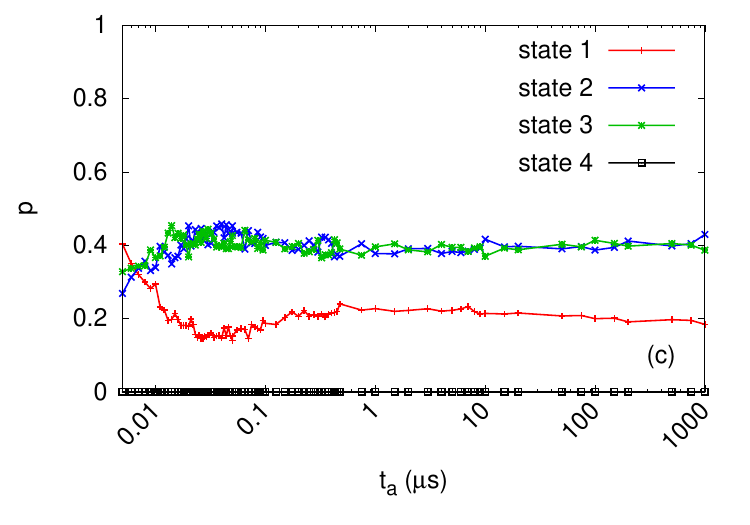}
         \put(-51, 102){\colorbox{white}{\makebox(12,2){\footnotesize $\ket{\uparrow\uparrow}$}}}
         \put(-51, 94){\colorbox{white}{\makebox(12,2){\footnotesize $\ket{\uparrow\downarrow}$}}}
         \put(-51, 86){\colorbox{white}{\makebox(12,2){\footnotesize $\ket{\downarrow\uparrow}$}}} 
         \put(-51, 78){\colorbox{white}{\makebox(12,2){\footnotesize $\ket{\downarrow\downarrow}$}}}
     \end{minipage}
     \hfill
     \caption{(Color online) D-Wave data obtained using the fast annealing protocol for a 2-spin problem instances with 
     (a) 2S1, (b) 2S2, and (c) 2S3.}
    \label{fig:motiv_fast_2spin}
\end{figure*}
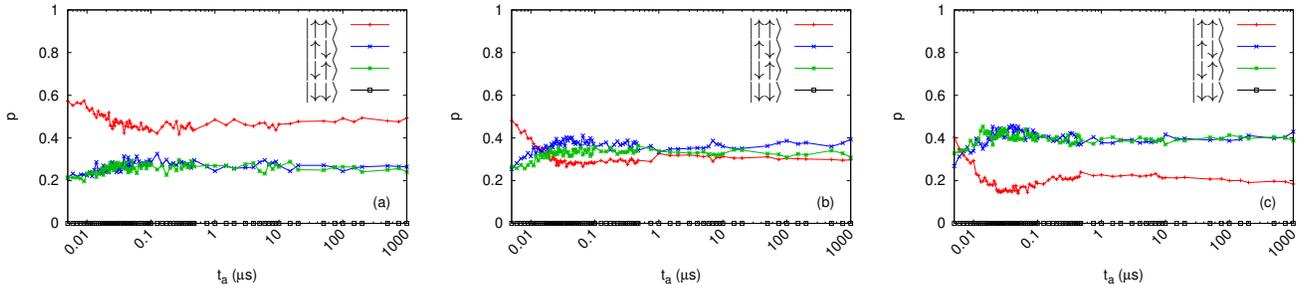

Moving next to the case of the 2-spin problems, Fig.~\ref{fig:motiv_fast_2spin} shows the D-Wave data for the three 2-spin instances. In this case, although we do not observe an initially increasing ground state probability with the annealing time (except for 2S3 in panel (c)), the initial trend of the probabilities is significantly different than those corresponding to standard annealing. Another important difference to the latter is the absence of the ``dips-and-bumps" (around $0.5~\mu s \leq t_a \leq 1.2~\mu s$) in this case.

\section{Simulation results}
\label{sec:simulation}
After analyzing the standard and fast annealing results obtained from the D-Wave annealers, we now focus on modeling these outcomes through appropriate simulations. Specifically, we employ Bloch equation simulations for 1-spin problems and utilize Lindblad and Markovian master equation simulations for 2-spin problems. Additionally, in Appendix~\ref{app:spinbath}, we present an alternative approach based on a spin system interacting with a heat bath, along with the corresponding results.
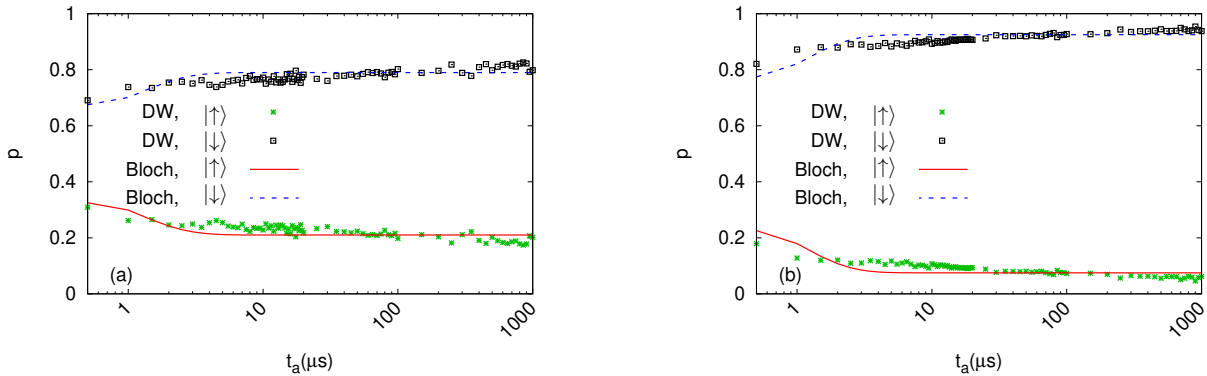
\begin{figure*}
     \begin{minipage}{0.49\textwidth}
         \centering
         \includegraphics[scale=0.6]{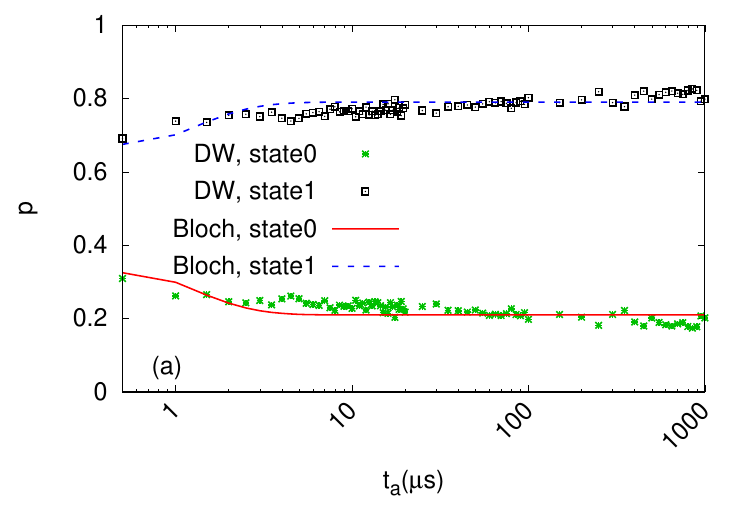}
         \put(-146, 103){\colorbox{white}{\makebox(20,6){\footnotesize $\ket{\uparrow}$}}}
         \put(-146, 93){\colorbox{white}{\makebox(20,6){\footnotesize $\ket{\downarrow}$}}}
         \put(-146, 83){\colorbox{white}{\makebox(20,6){\footnotesize $\ket{\uparrow}$}}} 
         \put(-146, 73){\colorbox{white}{\makebox(20,6){\footnotesize $\ket{\downarrow}$}}}
     \end{minipage}
     \hfill
     \begin{minipage}{0.49\textwidth}
         \centering
         \includegraphics[scale=0.6]{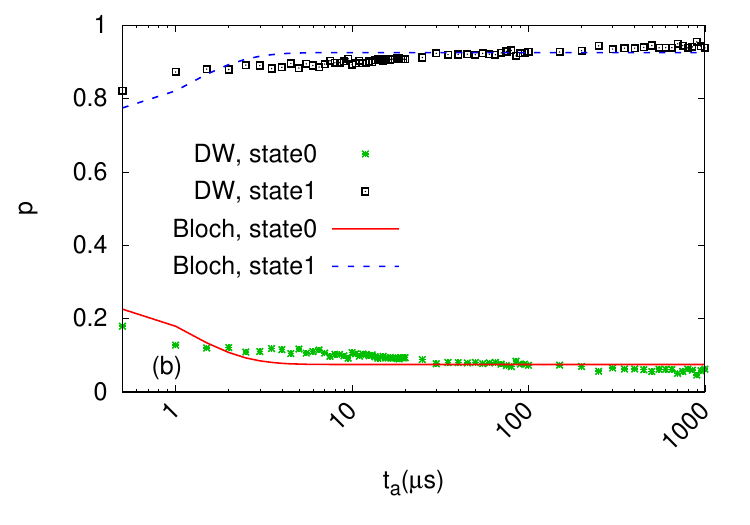}
         \put(-146, 103){\colorbox{white}{\makebox(20,6){\footnotesize $\ket{\uparrow}$}}}
         \put(-146, 93){\colorbox{white}{\makebox(20,6){\footnotesize $\ket{\downarrow}$}}}
         \put(-146, 83){\colorbox{white}{\makebox(20,6){\footnotesize $\ket{\uparrow}$}}} 
         \put(-146, 73){\colorbox{white}{\makebox(20,6){\footnotesize $\ket{\downarrow}$}}}
     \end{minipage} 
     \hfill
     \caption{(Color online) Comparison of Bloch equations simulation results with those from the D-Wave annealer for (a) $h_1=0.1$ and (b) $h_1=0.2$. For (a) $T_1 = 500$~ns, $T_2 = 125$~ns, and $M_0 =-0.58$ while for (b) $T_1=500$~ns, $T_2=125$~ns, and $M_0=-0.85$. In all figures with simulation results, the data points obtained from our simulations are, for improved legibility, represented by lines.}
    \label{fig:sim_1spin}
\end{figure*}

\subsection{Standard quantum annealing}
The qualitative agreement between the results from the annealer and the equilibrium probabilities observed so far makes it evident that the ideal quantum annealing simulations cannot reproduce the data obtained from the D-Wave annealers. Therefore, as for reverse annealing in Ref.~\cite{reverseannealing}, we use the Gorini–Kossakowski–Sudarshan–Lindblad (GKSL) master equation, which approximates the Schr\"odinger dynamics of a density matrix for a system interacting with an environment~\cite{gorini1976completely,Lindblad1976,breuer2002theory}.

The reduced master equation in Lindblad diagonal form reads
\begin{align}
    \frac{d \rho(t)}{dt} &= -\frac{i}{\hbar} [H(t), \rho(t)] \nonumber\\
    &+ \frac{1}{2} \sum_j \gamma_j (2L_j\rho(t)L_j^\dagger - L_j^\dagger L_j \rho(t) - \rho(t) L_j^\dagger L_j ),
    \label{eq:lindblad}
\end{align}
where $\rho(t)$ is the density matrix of the system and $\gamma_j \geq 0 $ are the dissipation rates corresponding to the operators $L_j$. In general, the operators $L_j$ are linear combinations of the matrices that form a basis for the matrices operating on the Hilbert space of the system \cite{breuer2002theory}. 

\subsubsection{1-spin problems: Bloch equations}
\label{sec:sim_1spin}
In general, the Hamiltonian for a system of one spin-1/2 object reads
\begin{equation}
    H = -\frac{1}{2}\mathbf{B} \cdot \boldsymbol{\sigma},
    \label{eq:1spinHamil}
\end{equation}
where \textbf{B} is the applied magnetic field and $\boldsymbol{\sigma}=(\sigma^x,\sigma^y,\sigma^z)$ are the Pauli matrices. Choosing dissipation operators $L_1 = \sigma^+=(\sigma^x+i\sigma^y)/2$, $L_2=\sigma^-=(\sigma^x-i\sigma^y)/2$ and $L_3=\sigma^z$, 

we find that the Lindblad master equation becomes equivalent to the Bloch equations~\cite{reverseannealing}, given by
\begin{align}
           {\frac {dS^{x}(t)}{dt}}&=S^{y}(t)B^{z}(t)-S^{z}(t)B^{y}(t)-{\frac {S^{x}(t)}{T_{2}}}\nonumber\\
           {\frac {dS^{y}(t)}{dt}}&=S^{z}(t)B^{x}(t)-S^{x}(t)B^{z}(t)-{\frac {S^{y}(t)}{T_{2}}}\nonumber\\
           {\frac {dS^{z}(t)}{dt}}&=S^{x}(t)B^{y}(t)-S^{y}(t)B^{x}(t))-{\frac {S^{z}(t)-M_{0}}{T_{1}}},
           \label{eq:bloch}
\end{align}
with $T_2 = 2/(\gamma_1 + \gamma_2 + 4\gamma_3)$, $T_1=1/(\gamma_1+\gamma_2)$, $M_0=(\gamma_1-\gamma_2)/(\gamma_1+\gamma_2)$ denoting the transverse and longitudinal relaxation time and the equilibrium magnetization, respectively. Furthermore, this choice of the dissipation operators yields $T_2 \leq 2T_1$ \cite{reverseannealing}.

In Fig.~\ref{fig:sim_1spin}, we show the numerically obtained results for the 1-spin problem with $h_1=0.1$ and $h_1=0.2$ corresponding to $T_1=500$~ns, $T_2=125$~ns, and $M_0=0.8~(0.85)$ for $h_1=0.1~(0.2)$. The close agreement of these results with those obtained from the annealer clearly shows that the Bloch equations with appropriate choices for $T_1$, $T_2$, and $M_0$ can reproduce the D-Wave results rather well. 

\subsubsection{2-spin problems: Lindblad master equation}
\label{sec:sim_2spin}

\begin{figure*}
     \begin{minipage}{0.33\textwidth}
         \centering
         \includegraphics[width=\textwidth]{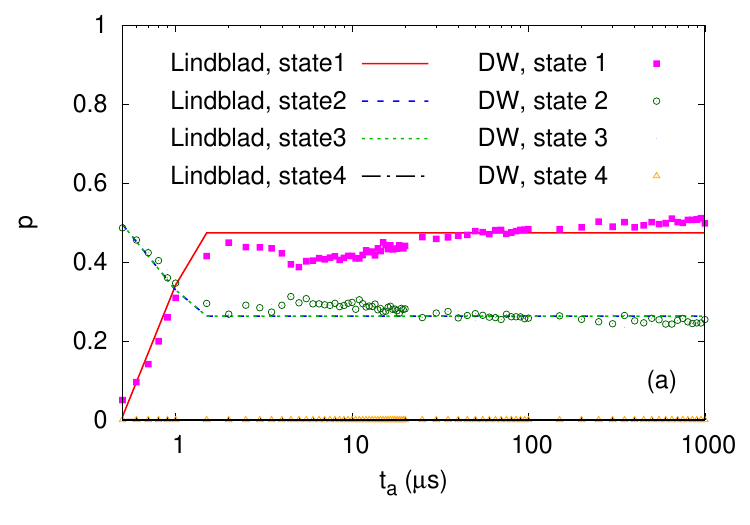}
         \put(-106.5, 100){\colorbox{white}{\makebox(12,2){\footnotesize $\ket{\uparrow\uparrow}$}}}
         \put(-106.5, 92){\colorbox{white}{\makebox(12,2){\footnotesize $\ket{\uparrow\downarrow}$}}}
         \put(-106.5, 84){\colorbox{white}{\makebox(12,2){\footnotesize $\ket{\downarrow\uparrow}$}}} 
         \put(-106.5, 75){\colorbox{white}{\makebox(12,2){\footnotesize $\ket{\downarrow\downarrow}$}}}
         \put(-50.5, 100){\colorbox{white}{\makebox(14,4){\footnotesize $\ket{\uparrow\uparrow}$}}}
         \put(-50.5, 92){\colorbox{white}{\makebox(14,2){\footnotesize $\ket{\uparrow\downarrow}$}}}
         \put(-50.5, 84){\colorbox{white}{\makebox(14,2){\footnotesize $\ket{\downarrow\uparrow}$}}} 
         \put(-50.5, 75){\colorbox{white}{\makebox(14,2){\footnotesize $\ket{\downarrow\downarrow}$}}}
     \end{minipage}
     \hfill
     \begin{minipage}{0.33\textwidth}
         \centering
         \includegraphics[width=\textwidth]{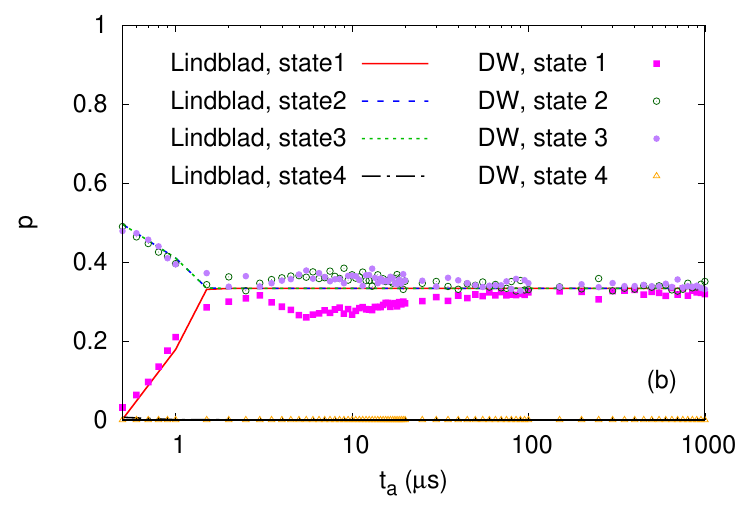}
         \put(-106.5, 100){\colorbox{white}{\makebox(12,2){\footnotesize $\ket{\uparrow\uparrow}$}}}
         \put(-106.5, 92){\colorbox{white}{\makebox(12,2){\footnotesize $\ket{\uparrow\downarrow}$}}}
         \put(-106.5, 84){\colorbox{white}{\makebox(12,2){\footnotesize $\ket{\downarrow\uparrow}$}}} 
         \put(-106.5, 75){\colorbox{white}{\makebox(12,2){\footnotesize $\ket{\downarrow\downarrow}$}}}
         \put(-50.5, 100){\colorbox{white}{\makebox(14,2){\footnotesize $\ket{\uparrow\uparrow}$}}}
         \put(-50.5, 92){\colorbox{white}{\makebox(14,2){\footnotesize $\ket{\uparrow\downarrow}$}}}
         \put(-50.5, 84){\colorbox{white}{\makebox(14,2){\footnotesize $\ket{\downarrow\uparrow}$}}} 
         \put(-50.5, 75){\colorbox{white}{\makebox(14,2){\footnotesize $\ket{\downarrow\downarrow}$}}}
     \end{minipage}
        \begin{minipage}{0.33\textwidth}
         \centering
         \includegraphics[width=\textwidth]{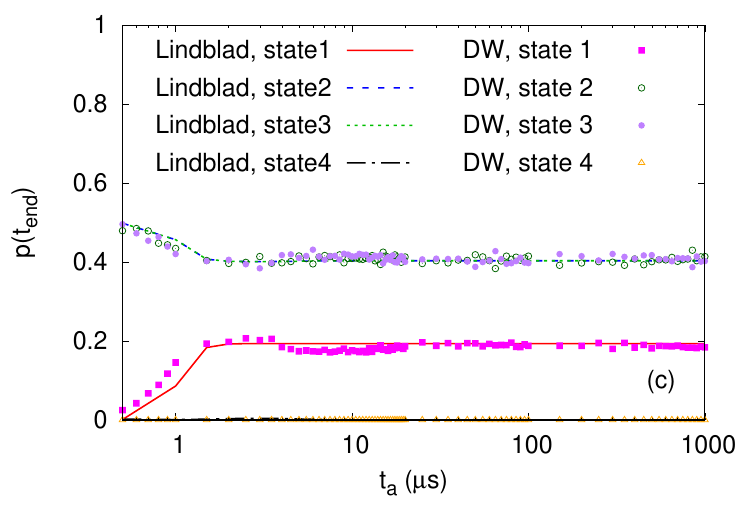}
         \put(-106.5, 100){\colorbox{white}{\makebox(12,2){\footnotesize $\ket{\uparrow\uparrow}$}}}
         \put(-106.5, 92){\colorbox{white}{\makebox(12,2){\footnotesize $\ket{\uparrow\downarrow}$}}}
         \put(-106.5, 84){\colorbox{white}{\makebox(12,2){\footnotesize $\ket{\downarrow\uparrow}$}}} 
         \put(-106.5, 75){\colorbox{white}{\makebox(12,2){\footnotesize $\ket{\downarrow\downarrow}$}}}
         \put(-50.5, 100){\colorbox{white}{\makebox(14,2){\footnotesize $\ket{\uparrow\uparrow}$}}}
         \put(-50.5, 92){\colorbox{white}{\makebox(14,2){\footnotesize $\ket{\uparrow\downarrow}$}}}
         \put(-50.5, 84){\colorbox{white}{\makebox(14,2){\footnotesize $\ket{\downarrow\uparrow}$}}} 
         \put(-50.5, 75){\colorbox{white}{\makebox(14,2){\footnotesize $\ket{\downarrow\downarrow}$}}}
     \end{minipage}
     \hfill
     \caption{(Color online) Comparison of the D-Wave with that from Lindblad master equation simulation for 2-spin instance (a) 2S1, (b) 2S2, and (c) 2S3, with  (a)~$T=35$~mK and $c=0.01$, (b)~$T=28$~mK and $c=0.003$, and  (c)~$T=28$~mK and $c=0.001$.}
    \label{fig:sim_2spin}
\end{figure*}

\begin{figure*}
          \begin{minipage}{0.33\textwidth}
         \centering
         \includegraphics[width=\textwidth]{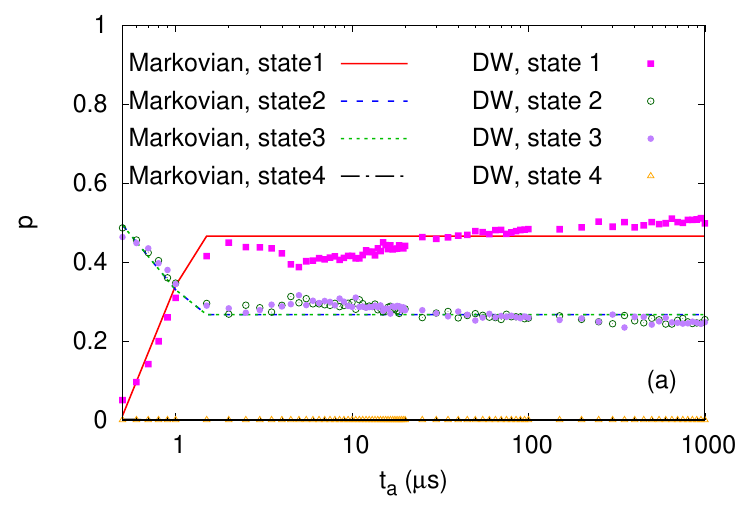}
         \put(-108.5, 100){\colorbox{white}{\makebox(12,2){\footnotesize $\ket{\uparrow\uparrow}$}}}
         \put(-108.5, 92){\colorbox{white}{\makebox(12,2){\footnotesize $\ket{\uparrow\downarrow}$}}}
         \put(-108.5, 84){\colorbox{white}{\makebox(12,2){\footnotesize $\ket{\downarrow\uparrow}$}}} 
         \put(-108.5, 75){\colorbox{white}{\makebox(12,2){\footnotesize $\ket{\downarrow\downarrow}$}}}
         \put(-50.5, 100){\colorbox{white}{\makebox(14,4){\footnotesize $\ket{\uparrow\uparrow}$}}}
         \put(-50.5, 92){\colorbox{white}{\makebox(14,2){\footnotesize $\ket{\uparrow\downarrow}$}}}
         \put(-50.5, 84){\colorbox{white}{\makebox(14,2){\footnotesize $\ket{\downarrow\uparrow}$}}} 
         \put(-50.5, 75){\colorbox{white}{\makebox(14,2){\footnotesize $\ket{\downarrow\downarrow}$}}}
     \end{minipage}
     \hfill
     \begin{minipage}{0.33\textwidth}
         \centering
         \includegraphics[width=\textwidth]{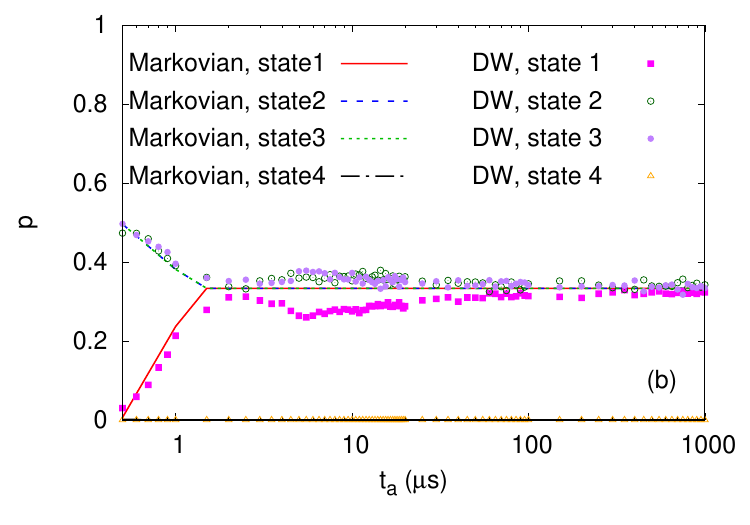}
         \put(-108.5, 100){\colorbox{white}{\makebox(12,2){\footnotesize $\ket{\uparrow\uparrow}$}}}
         \put(-108.5, 92){\colorbox{white}{\makebox(12,2){\footnotesize $\ket{\uparrow\downarrow}$}}}
         \put(-108.5, 84){\colorbox{white}{\makebox(12,2){\footnotesize $\ket{\downarrow\uparrow}$}}} 
         \put(-108.5, 75){\colorbox{white}{\makebox(12,2){\footnotesize $\ket{\downarrow\downarrow}$}}}
         \put(-50.5, 100){\colorbox{white}{\makebox(14,4){\footnotesize $\ket{\uparrow\uparrow}$}}}
         \put(-50.5, 92){\colorbox{white}{\makebox(14,2){\footnotesize $\ket{\uparrow\downarrow}$}}}
         \put(-50.5, 84){\colorbox{white}{\makebox(14,2){\footnotesize $\ket{\downarrow\uparrow}$}}} 
         \put(-50.5, 75){\colorbox{white}{\makebox(14,2){\footnotesize $\ket{\downarrow\downarrow}$}}}
     \end{minipage}
     \hfill
     \begin{minipage}{0.33\textwidth}
         \centering
         \includegraphics[width=\textwidth]{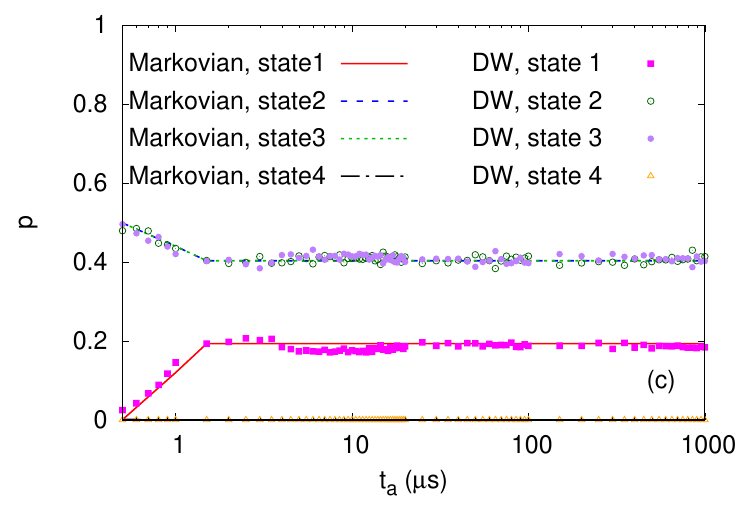}
         \put(-108.5, 100){\colorbox{white}{\makebox(12,2){\footnotesize $\ket{\uparrow\uparrow}$}}}
         \put(-108.5, 92){\colorbox{white}{\makebox(12,2){\footnotesize $\ket{\uparrow\downarrow}$}}}
         \put(-108.5, 84){\colorbox{white}{\makebox(12,2){\footnotesize $\ket{\downarrow\uparrow}$}}} 
         \put(-108.5, 75){\colorbox{white}{\makebox(12,2){\footnotesize $\ket{\downarrow\downarrow}$}}}
         \put(-50.5, 100){\colorbox{white}{\makebox(14,4){\footnotesize $\ket{\uparrow\uparrow}$}}}
         \put(-50.5, 92){\colorbox{white}{\makebox(14,2){\footnotesize $\ket{\uparrow\downarrow}$}}}
         \put(-50.5, 84){\colorbox{white}{\makebox(14,2){\footnotesize $\ket{\downarrow\uparrow}$}}} 
         \put(-50.5, 75){\colorbox{white}{\makebox(14,2){\footnotesize $\ket{\downarrow\downarrow}$}}}
     \end{minipage} 
     \hfill
     \caption{(Color online) Comparison of the D-Wave with that from Markovian master equation simulation for 2-spin instance (a) 2S1, (b) 2S2, and (c) 2S3, with  (a)~$T=35$~mK and $c=0.01$, (b)~$T=28$~mK and $c=0.003$, and  (c)~$T=28$~mK and $c=0.001$.}
    \label{fig:sim_pdotwp}
\end{figure*}
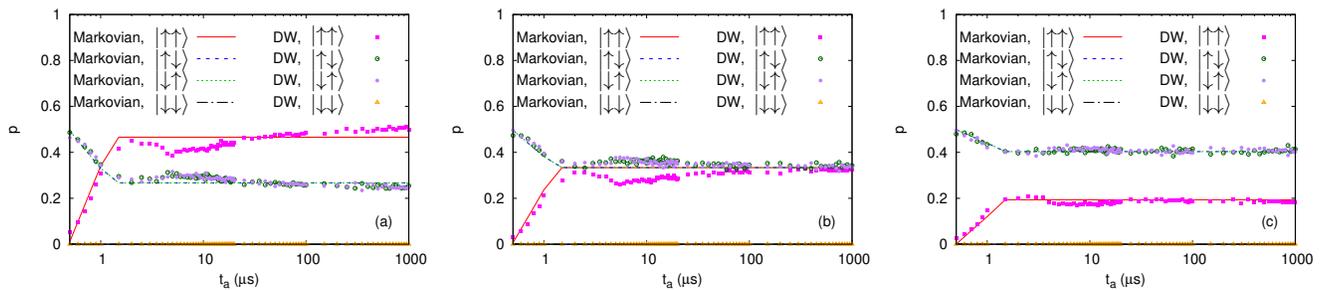
Next, we move on to the 2-spin instances. For consistency, we use the same set of seven dissipation operators in Eq.~(\ref{eq:lindblad}) as those used to study the reverse annealing protocol ~\cite{reverseannealing}.

The D-Wave data for these problems are shown in Fig.~\ref{fig:motiv_2spin}. As the annealing times chosen are much longer than those required for an adiabatic evolution of the initial uniform superposition state to the ground state of the problem Hamiltonian, it is not possible to reproduce the initial ``dip-and-bump" (initial $p_{\uparrow\uparrow} \approx 0$) through ideal quantum annealing simulations. Furthermore, unless somehow the stationary state for these problems itself changes for these problems as a function of the annealing time, within the Lindblad master equation picture, one cannot justify the probability of state $\ket{\uparrow\uparrow}$ being close to zero for the initial values of $t_a$ (see Fig.~\ref{fig:2spin_woonset} in Appendix~\ref{app:wo_onset} for an attempt to capture the behavior of resulting $p(t_a)$ for instance 2S2 with Lindblad master equation). However, these ``dips-and-bumps" are absent in the corresponding results of fast annealing, i.e., in the absence of magnetic fields $h_i$'s. This hints at a significant effect of non-zero magnetic fields during the initial phase of the annealing process.

On the physical level, the D-Wave QPU is made up of superconducting flux qubits, and the effective Hamiltonian for this system is given by~\cite{Harris2010}
\begin{align}
    H_{phys} &= -\frac{1}{2} \sum_i \Delta_q(\Phi_{CCJJ}(s))\sigma_x^i -2 h_i|I_p(\Phi_{CCJJ}(s))|\Phi_x^i(s) \sigma_z^i
    \nonumber\\ &+ \sum_{i>j} J_{ij} M_{AFM} I_p(\Phi_{CCJJ}(s))^2\sigma_z^i\sigma_z^j,
    \label{eq:DWHamilphys}
\end{align}
where $\Delta_q$ is the energy difference between the two eigenstates of the qubit ($\ket{0}\pm\ket{1}/\sqrt{2}$) with zero external flux, $I_p$ represents the magnitude of the current flowing in the body of the qubit loop, $M_{AFM}$ is the maximum mutual inductance generated by the couplers between the qubits, $\Phi_x^i(s)$ is an external flux applied to the qubits, and $\Phi_{CCJJ}(s)$ is an external flux applied to every qubit’s compound Josephson-junction structures to change the potential energy shape of the qubit. 

To map Hamiltonian Eq.~(\ref{eq:DWHamilphys}) on to Eq.~(\ref{eq:DWHamil}), $\Phi_i^x$ needs to be set equal to $M_{AFM}|I_p|$ in the standard annealing protocol, yielding
\begin{align}
    A(s) = \Delta_q (\Phi_{CCJJ}(s)) \nonumber\\
    B(s) = 2M_{AFM}|Ip(\Phi_{CCJJ}(s))|^2.
\end{align}
However, as per~\cite{Dwave}, for the very fast ramps in $\Phi_{CCJJ}(s)$ it is difficult to maintain the relative energy ratio between $h_i$ and $J_{ij}$ terms (by adjusting $\Phi_i^x = M_{AFM}|I_p|$). This knowledge, in combination with the D-Wave data showing the ``dips-and-bumps" and the fact that non-zero values of $h_i$ are not permitted on D-Wave for fast annealing 
 motivates us to the following. 
 
\textbf{Conjecture:} For the initial values of the annealing times, $\Phi_i^x(s)$ starts from a value close to zero and only gradually catches on to match the value set by the product $M_{AFM}|I_p|$. 

With this conjecture, it already seems possible to generate the ``dips-and-bumps" for the three instances of the 2-spin problems. If $\Phi_i^x$ starts from a value close to zero, the D-Wave annealer effectively implements a problem with $h_1=h_2=0$. For these altered problems, states $\ket{\uparrow\downarrow}$ and $\ket{\downarrow\uparrow}$ are the two degenerated ground states, while the other two states are the first excited states. 

To implement this idea, we adopt two separate annealing schemes for the linear and quadratic parts of the problem Hamiltonian, i.e.,
\begin{eqnarray}
    \frac{\widehat{H}(t)}{\hbar} &=&\frac{\pi}{h}\left[B'(s)\sum_i \sigma_i^z + B(s) \sum_{i>j} J_{i,j}\sigma_i^z\sigma_j^z\right]\;,\label{eq:h_hat}\\
    \frac{H(t)}{\hbar} &=&-\frac{\pi A(s)}{h} \sum_i \sigma_i^x 
    + \widehat{H}(t)
    \;,\label{eq:h_mod}
\end{eqnarray}
where 
\begin{widetext}
    \begin{equation}
    B'(s) = B'(t/t_a) = \begin{cases}
        0, &t \leq t_{onset}^{start}\\
        \sin\left(\frac{\pi}{2}\frac{(t-t_{onset}^{start})}{t_{onset}^{end}-t_{onset}^{start}}\right)B(s), &t_{onset}^{start} \leq t \leq t_{onset}^{end}\\
        B(s), &t > t_{onset}^{end},
    \end{cases}
    \label{eq:b'}
    \end{equation}
\end{widetext}
and $t_{onset}^{start}$ and $t_{onset}^{end}$ control the start and end of times for this ``out-of-synchronization" feature, respectively. However, modifying the annealing schedule is by itself insufficient for reproducing the ``dips-and-bumps". It is also necessary to let the dissipation rates depend on time.

The dissipation rates for the Lindblad master equation are related to the stationary state probabilities through~\cite{reverseannealing}
\begin{equation}
    \gamma_2 p_{\uparrow\uparrow} = \gamma_1 p_{\downarrow\downarrow},\; \gamma_5 p_{\uparrow\uparrow} = \gamma_4 p_{\downarrow\uparrow},\; \gamma_7 p_{\uparrow\uparrow} = \gamma_6 p_{\uparrow\downarrow}.
    \label{eq:pdissiprel}
\end{equation}
In the absence of the ``dips-and-bumps", $t_{onset}^{end}=0$ and we use the Hamiltonian $\widehat{H}(t_a)/\hbar$ to determine probabilities according to the Gibbs distribution (Eq.~(\ref{eq:equil})). The dissipation rates are then obtained from Eq.~(\ref{eq:pdissiprel}) such that the probabilities match those obtained from the D-Wave annealers for an appropriate value of $\beta$~\cite{reverseannealing}. However, as per the conjecture mentioned above, the Hamiltonian, according to Eq.~(\ref{eq:h_hat}), depends on the total annealing time $t_a$ and parameters $t_{onset}^{start}$ and $t_{onset}^{end}$. Consequently, following the same approach, according to Eq.~(\ref{eq:b'}), the dissipation rates for the Lindblad master equation now change as a function of time $t$ and are, as in the case of the absence of ``dips-and-bumps", obtained from the Gibbs distribution with Hamiltonian Eq.~(\ref{eq:h_hat}). 
To this end, we choose a value of $\beta$ that yields similar probabilities as those obtained from the D-Wave annealers at $t_a=1950~\mu$s, set $\gamma_1 = \gamma_3 = \gamma_4 = \gamma_6=c$ where $c$ is a fitting parameter, and determine $\gamma_2$, $\gamma_5$, and $\gamma_7$ using Eq.~(\ref{eq:pdissiprel}).

In Fig.~\ref{fig:sim_2spin}, we show the results for the 2-spin instances by setting $t_{onset}^{start}=0~\mu$s and $t_{onset}^{end}=1.2~\mu$s, and choosing the same dissipation operators as in the case for the reverse annealing simulations discussed in Ref.~\cite{reverseannealing}. The qualitative agreement between the numerical results and the D-Wave data increases our confidence in the conjecture. Although it might be possible to obtain a better match of the simulation results by tailoring the $B'(t)$ annealing function to the D-Wave data, the focus of the current paper is to only capture the qualitative behavior of the data, not to tweak parameters for the best match.

\begin{figure}[H]
    \centering
    \includegraphics[scale=0.6]{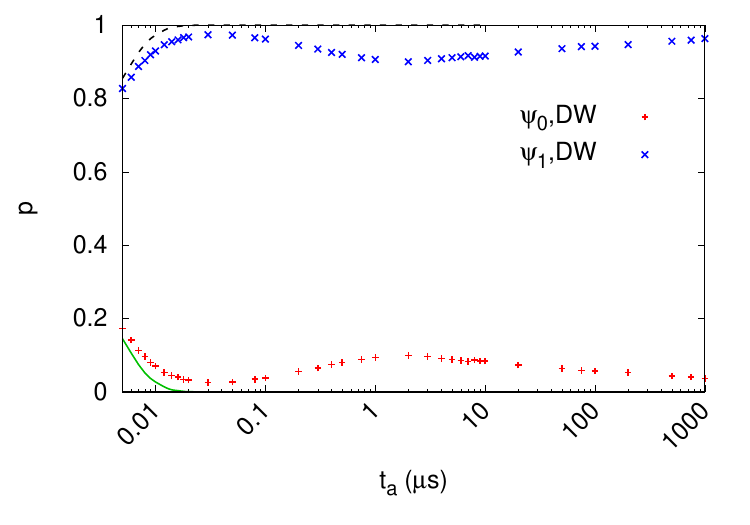}
    \put(-67, 114){\colorbox{white}{\makebox(21,6){\footnotesize $\ket{\uparrow}$}}}
    \put(-67, 104){\colorbox{white}{\makebox(21,6){\footnotesize $\ket{\downarrow}$}}}
    \caption{(Color online) Comparison of D-Wave data using the fast annealing protocol for 1-spin problem with $h_1=0.25$ with ideal quantum annealing simulations.}
    \label{fig:fast_sim_1spin}
\end{figure}
\subsubsection{Larger problems: Markovian master equation}
\label{sec:pdotwp}

In our reverse annealing study~\cite{reverseannealing}, we noted that it is possible to reproduce the D-Wave behavior using the remarkably simple procedure of ignoring the non-diagonal elements of the density matrix $\rho(t)$ and setting $A(s)=0$~\cite{reverseannealing}. 
Likewise, in the case at hand, except for the initial part of the annealing process, the simulations of the Lindblad master equation show that the absolute values of the non-diagonal elements of the density matrix quickly tend to be rather small. 

In the present section, we investigate whether ignoring the non-diagonal elements of $\rho(t)$ can still reproduce the results in the regime of interest. This approach amounts to solving
\begin{equation}
    \frac{d P(t)}{dt} = W(t)P(0),
    \label{eq:markov_diff}
\end{equation}
where $P(t) = (p_1(t), . . . , p_4 (t))^T$ is a vector of non-negative elements which sum to one and $W(t)$ is a real-valued matrix, and 
\begin{equation}
    W(t) = \begin{pmatrix}
        -\gamma_2-\gamma_5-\gamma_7 & \gamma_4 & \gamma_6 & \gamma_1 \\
        \gamma_5 & -\gamma_4 & 0 & 0 \\ 
        \gamma_7 & 0 & -\gamma_6 & 0 \\ 
        \gamma_2 & 0 & 0 & -\gamma_1
    \end{pmatrix},
    \label{eq:wmatrix}
\end{equation}
where $\gamma_i$'s are time-dependent and are chosen in the same way as for the simulations of the Lindblad master equation. As the columns of $W$ (Eq.~(\ref{eq:wmatrix})) add to zero, it follows immediately that Eq.~(\ref{eq:markov_diff}) describes a Markov process.

In Fig.~\ref{fig:sim_pdotwp}, we show the results obtained from the Markovian master equation simulation in comparison to the data from D-Wave. As for the Lindblad simulations, we incorporate different annealing schedules for the $h_i$ and $J_{ij}$ terms in Eq.~\ref{eq:h_hat} and as for Fig.~\ref{fig:sim_2spin}, we choose $t_{onset}^{start}=0~\mu$s and $t_{onset}^{end}=1.2~\mu$s. The close agreement between these two demonstrates that it is possible to reproduce the D-Wave results by circumventing the problem of finding the appropriate dissipation operators even for the standard quantum annealing protocol, that is, by a simple (non-quantum) Markovian model.

\subsection{Fast annealing}
\label{sec:fast_ann_sim}

Next, we move on to the fast annealing simulations for the 1- and 2-spin problems. Although, as explained above, running these problems with the fast annealing feature on the D-Wave requires using an extra qubit for each qubit with a non-zero $h_i$, we carry out the numerical implementation in a straightforward way, i.e., without introducing additional qubits in our simulations. Furthermore, as the ``dips-and-bumps" observed in the D-Wave data obtained with the standard annealing protocol are absent when using the fast annealing protocol, one does not need to implement the different annealing schedules for the linear and quadratic terms of the problem Hamiltonian. This is in concert with the conjecture captured by Eq.~(\ref{eq:h_mod}). Moreover, as the D-Wave results for the 1- and 2-spin problems eventually tend to equilibrium values, an effect that has been demonstrated to be described well by the master equation simulations, the most interesting information that can be extracted using the fast annealing feature is a signature for quantum coherence. Therefore, in this section, we compare the D-Wave results with the ideal quantum annealing simulations.

\begin{figure*}
     \begin{minipage}{0.33\textwidth}
         \centering
         \includegraphics[width=\textwidth]{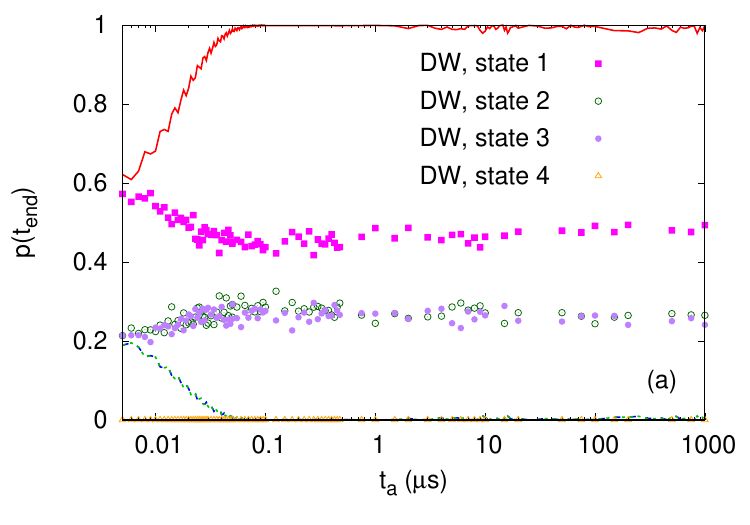}
         \put(-63, 100){\colorbox{white}{\makebox(15,2){\footnotesize $\ket{\uparrow\uparrow}$}}}
         \put(-63, 92){\colorbox{white}{\makebox(15,2){\footnotesize $\ket{\uparrow\downarrow}$}}}
         \put(-63, 84){\colorbox{white}{\makebox(15,2){\footnotesize $\ket{\downarrow\uparrow}$}}} 
         \put(-63, 75){\colorbox{white}{\makebox(15,2){\footnotesize $\ket{\downarrow\downarrow}$}}}
     \end{minipage}
     \hfill
     \begin{minipage}{0.33\textwidth}
         \centering
         \includegraphics[width=\textwidth]{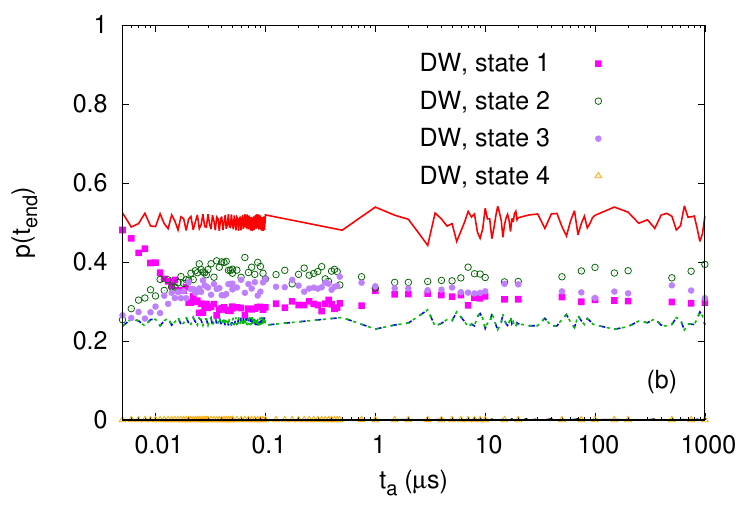}
         \put(-63, 100){\colorbox{white}{\makebox(15,2){\footnotesize $\ket{\uparrow\uparrow}$}}}
         \put(-63, 92){\colorbox{white}{\makebox(15,2){\footnotesize $\ket{\uparrow\downarrow}$}}}
         \put(-63, 84){\colorbox{white}{\makebox(15,2){\footnotesize $\ket{\downarrow\uparrow}$}}} 
         \put(-63, 75){\colorbox{white}{\makebox(15,2){\footnotesize $\ket{\downarrow\downarrow}$}}}
     \end{minipage}
        \begin{minipage}{0.33\textwidth}
         \centering
         \includegraphics[width=\textwidth]{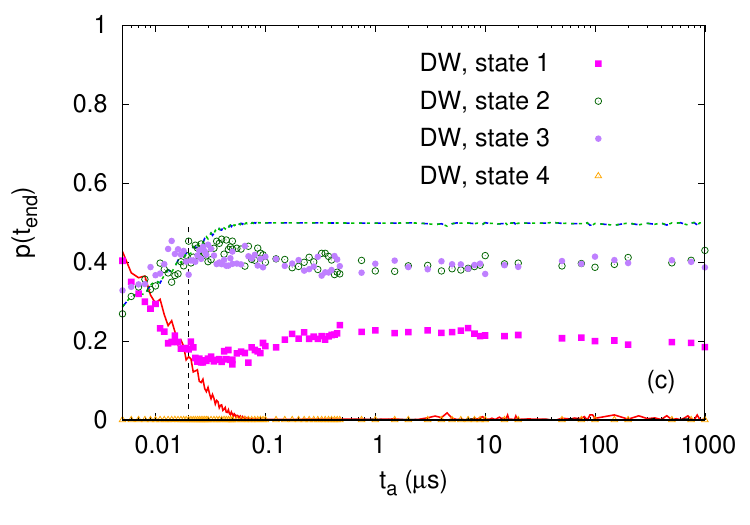}
         \put(-63, 100){\colorbox{white}{\makebox(15,2){\footnotesize $\ket{\uparrow\uparrow}$}}}
         \put(-63, 92){\colorbox{white}{\makebox(15,2){\footnotesize $\ket{\uparrow\downarrow}$}}}
         \put(-63, 84){\colorbox{white}{\makebox(15,2){\footnotesize $\ket{\downarrow\uparrow}$}}} 
         \put(-63, 75){\colorbox{white}{\makebox(15,2){\footnotesize $\ket{\downarrow\downarrow}$}}}
     \end{minipage}
     \hfill
     \caption{(Color online) Comparison of the D-Wave using the fast annealing protocol with ideal quantum annealing simulations for 2-spin instance (a) 2S1, (b) 2S2, and (c) 2S3. The dashed line in panel (c) corresponds to $t_a=20$~ns.}
    \label{fig:fast_sim_2spins}
\end{figure*}
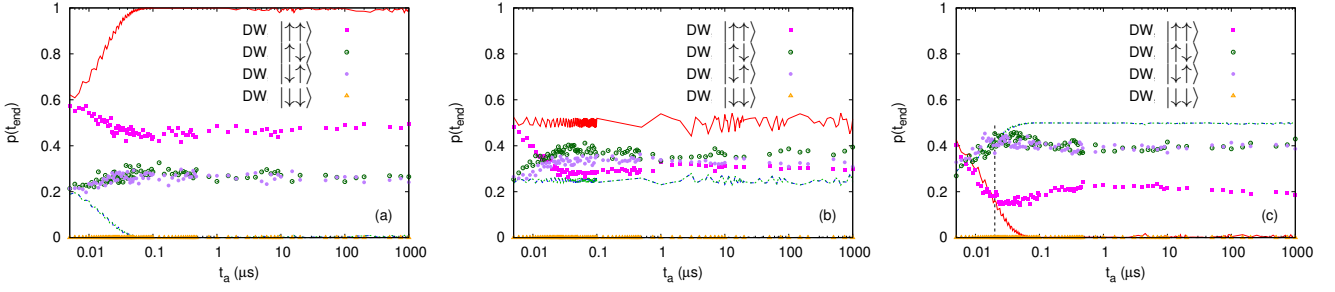

Shown in Figs.~\ref{fig:fast_sim_1spin} and \ref{fig:fast_sim_2spins} are the comparison of the ideal quantum annealing simulations with the D-Wave annealer data for 1- and 2-spin problems, respectively. From these figures, it can be noted that the numerical results match the D-Wave data closely only for the shortest annealing time permitted on the D-Wave systems, i.e., $t_a=5$~ns. Beyond this value, the D-Wave data starts to deviate from the coherent simulation results, suggesting that only for times less than $t_a = 5$~ns, can we imagine quantum coherence to play a role in the dynamics of the Advantage\_5.4 machine. Note that other systematic biases/imperfections in the D-Wave systems, e.g., the systematic decrease in the probabilities corresponding to $\ket{\uparrow\uparrow}$ state in Fig.~\ref{fig:fast_sim_2spins}(c) can be misleading and be mistaken for a sign of longer coherence times ($t_a~\approx 20$~ns) for specific problems, as in the case of problem 2S3. 

For completeness, in Appendix~\ref{app:additional_fast}, we show the fast annealing results for additional 2- and 3-spin problems that have zero magnetic fields $h_i$'s.

\section{Conclusion}
\label{sec:conclusion}

Focusing on simple 1- and 2-spin problems, we demonstrated that for sufficiently long annealing times, both the standard and fast annealing protocols on D-Wave annealers sample states with frequencies approaching thermal equilibrium, as has also been shown for reverse annealing \cite{reverseannealing}.

By numerically solving the GKSL master equation with appropriately chosen dissipation operators and rates, we successfully reproduced the D-Wave data obtained from the standard forward annealing protocol. In most cases, this was achievable using the annealing schedule provided by D-Wave. However, in certain instances, replicating the results required modifications to the annealing schedule. Additionally, the empirical data aligned well with a classical Markovian process, suggesting that non-quantum models can effectively describe the system's behavior.

A comparison between empirical data from fast annealing and ideal quantum annealing simulations revealed no clear evidence of quantum behavior beyond annealing times of about $5$~ns.

Our study adds to the understanding of the physical mechanisms governing D-Wave annealers and underscores the need to explore strategies that leverage these insights for improved utilization in optimization problems. Importantly, our findings do not diminish the potential of D-Wave annealers as optimization solvers but rather provide a deeper understanding of the underlying physical processes driving their operation. 

If, as usual, the solution to the optimization problem is encoded in the ground state of the Ising Hamiltonian, one possible approach to leverage D-Wave systems to solve the problem efficiently is to lower the QPU temperature. Ref.~\cite{Hen_tempscaling_2017} shows that a finite temperature annealer requires logarithmically, or possibly polynomially, decreasing temperatures for solving problems with increasing sizes. An alternative strategy is to encode the solution close to the mean energy of the problem Hamiltonian, a strategy that requires further investigation. One possible direction to explore is the  quantum annealing correction Hamiltonian~\cite{Pudenz_2014} which effectively encodes the solution to the optimization problem not solely in the ground state but also within a small block of low-energy excited states.

\section{Acknowledgements}
The authors gratefully acknowledge the Gauss Centre for Supercomputing e.V. (www.gauss-centre.eu) for funding this project by providing computing time through the John von Neumann Institute for Computing (NIC) on the GCS Supercomputer JUWELS~\cite{JUWELS} at the Jülich Supercomputing Centre (JSC). V.M. acknowledges support from the project JUNIQ funded by the German Federal Ministry of Education and Research (BMBF) and the Ministry of Culture and Science of the State of North Rhine-Westphalia (MKW-NRW) and from the project EPIQ funded by MKW-NRW.

\appendix

\section{Annealing schedules for standard and fast annealing}
\label{app:sched}

\begin{figure}
    \centering
    \includegraphics[scale=0.5]{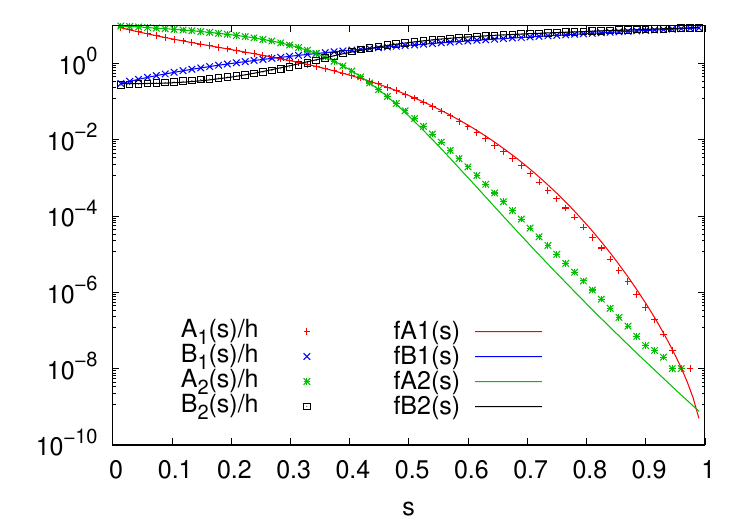}
    \caption{Annealing schemes for the standard ($A_1(s)/h$, $B_1(s)/h$) and the fast ($A_2(s)/h$, $B_2(s)/h$) annealing protocols for the 
    D-Wave Adv\_5.4 system, and the corresponding fits ($f_{A_1}(s)/h$, $f_{B_1}(s)/h$) and ($f_{A_2}(s)/h$, $f_{B_2}(s)/h$), see Eqs.~(\ref{eq:scheme1}),(\ref{eq:scheme2}).}
    \label{fig:annealingscheme}
\end{figure}

The D-Wave data for the annealing scheme are provided as tabulated values of $A(s)/h$ and $B(s)/h$ (in GHz)~\cite{Dwave}. For our numerical simulation, we fit the functions
\begin{align}
    &f_{A_1}(s)/h = (1-s) \exp(a_1 + a_2s + a_3s^2 + a_4 s^3),\nonumber \\
    &f_{B_1}(s)/h = b_1 + b_2s + b_3s^2,
    \label{eq:scheme1}
\end{align}
for standard quantum annealing with $a_1=2.27$, $a_2=-8.22$, $a_3=16.14$, $a_4=-27.59$, $b_1=0.26$, $b_2=2.46$, and $b_3=5.86$, and the functions
\begin{align}
    f_{A_2}(s)/h =& \exp(f_0(s)(c_1+c_2s+c_3s^3+c_4s^4)\nonumber \\
    &+(1-f_0(s))(c_5+c_6s)), \nonumber\\
    f_{B_2}(s)/h =& \exp(d_1+d_2(1-s)\tanh(d_3s^{3/2})\nonumber\\
    &+d_4\tanh(d_5s^2)), \nonumber \\
    f_0(s) &= 0.5(1+\tanh(a_0(b_0-s))    
    \label{eq:scheme2}
\end{align}
for the fast annealing protocol with $c_1=2.15$, $c_2=-2.66$, $c_3=-35.29$, $c_4=143.48$, $c_5=8.99$, $c_6=-30.63$ $d_1=-1.21$, $d_2=-1.24$, $d_3=4.79$, $d_4=3.38$, $d_5=5.87$, $a_0=5.00$, and $b_0=0.40$.

\section{Extracting model parameters from D-Wave data}
\label{app:maxent}
It is well known that the parameters specified to the D-Wave systems are subject to inherent noise, which can impact the accuracy and reliability of the annealing process. An important question is whether, using the empirical data (frequencies corresponding to the four energy states of the 2-variable problems) from the D-Wave annealers, we can determine a model that best describes the data.

Application of the principle of maximum entropy to the D-Wave data, subject to constraints $\sum_i p_i = 1$ and $\langle E_k \rangle = \sum_i p_i E_k(x_i)$, $k=1,\dots, K$, where $p_i = p(x_i |\langle E_1 \rangle,\dots, \langle E_k \rangle Z)$ is the probability for the event $x_i$, the $\langle E_k \rangle$ are the mean values of the $E_k(x_i)$’s and $Z$ encodes all other facts, yields a probability distribution of form
    \begin{equation}
        p_i = \frac{e^{-\lambda_1 E_1(x_i)-\ldots-\lambda_K E_k(x_i)}}{\sum_i e^{-\lambda_1 E_1(x_i)-\ldots-\lambda_K E_k(x_i)}},
        \label{eq:app_maxent_1}
    \end{equation}
where $\lambda_0,\dots, \lambda_K$ are the Lagrange multipliers introduced to maximize the entropy, $S=-\sum_i p_i \ln p_i$, conditioned on the specified constraints. Obviously, Eq.~(\ref{eq:app_maxent_1}) has the same structure as Gibb's distribution (Eq.~(\ref{eq:equil})). As is well known, invoking the principle of maximum entropy to determine the probability distribution that describes the empirical data is equivalent to describing the empirical data by the probability distribution of equilibrium statistical mechanics~\cite{Jaynes1957a,Jaynes1957b}.

The problem parameters $\beta h_1$, $\beta h_2$, and $\beta J$ can be extracted by determining the Lagrange multipliers.
We use two methods based on two different sets of constraints.

\textbf{Method 1: }From the empirical frequencies $f_1$, $f_2$, $f_3$, and $f_4$ for energy levels corresponding to states $x_1=(S_1=+1,S_2=+1)$, $x_2=(S_1=+1,S_2=-1)$, $x_3=(S_1=-1,S_2=+1)$, and $x_4=(S_1=-1,S_2=-1)$, respectively, one can calculate the spin averages $\widehat{S}_1$, $\widehat{S}_2$, and the correlation $\widehat{S}_{12}$ using
\begin{align}
    \widehat{S}_1 = f_1+&f_2-f_3-f_4,\;\; \widehat{S}_2 = f_1-f_2+f_3-f_4 \nonumber \\
    &\widehat{S}_{12} = f_1-f_2-f_3+f_4.
    \label{eq:app_freq}
\end{align}
Considering the three averages in Eq.~(\ref{eq:app_freq}) as constraints for the maximum entropy, from Eq.~(\ref{eq:app_maxent_1}) it follows
\begin{equation}
    p_i = \frac{e^{-\lambda_1 S_1-\lambda_2 S_2-\lambda_3 S_{12}}}{\sum_{S_1,S_2=\pm 1} e^{-\lambda_1 S_1-\lambda_2 S_2-\lambda_3 S_{12}}},
    \label{eq:app_meth1_1}
\end{equation}
where the three Lagrange parameters have to be determined such that the three constraints
\begin{align}
    \widehat{S}_1 = \sum_{S_1,S_2=\pm 1} S_1 &p(S_1,S_2) ,\;\; \widehat{S}_2 = \sum_{S_1,S_2=\pm 1} S_2p(S_1,S_2) , \nonumber\\
    &\widehat{S}_{12} = \sum_{S_1,S_2=\pm 1}  S_{12}p(S_1,S_2) ,
    \label{eq:app_meth1_2}
\end{align}
are satisfied. These equations can be solved analytically for $\lambda_1$, $\lambda_2$, and $\lambda_3$, yielding

\begin{align}
    \lambda_1 = \frac{1}{4}\ln\frac{(a+b-c-1)(a-b+c-1)}{(a-b-c+1)(a+b+c+1)} \nonumber \\
    \lambda_2 = \frac{1}{4}\ln\frac{(a-b-c+1)(a+b-c-1)}{(a-b+c-1)(a+b+c+1)} \nonumber \\
    \lambda_3 = \frac{1}{4}\ln\frac{(a-b-c+1)(a-b+c-1)}{(a+b-c-1)(a+b+c+1)},
    \label{eq:app_meth1_3}
\end{align}
where $a=\widehat{S}_1$, $b=\widehat{S}_2$, and $c=\widehat{S}_{12}$. In terms of the frequencies, we have
\begin{align}
    \lambda_1 = \frac{1}{4}\ln\frac{f_2 f_4}{f_1 f_3} ,\;
    \lambda_2 = \frac{1}{4}\ln\frac{f_3 f_4}{f_1 f_2} ,\;
    \lambda_3 = \frac{1}{4}\ln\frac{f_2 f_3}{f_1 f_4}.
    \label{eq:app_meth1_4}
\end{align}
From Eq.~\ref{eq:app_meth1_4} it immediately follows that this method for extracting the parameters will fail if one of the four frequencies is zero. In practice, this is to be expected when the theoretical probability of sampling a state is very small.

Next, to determine the inverse temperature $\beta$ we use the parameters $h_1$, $h_2$, $J$ that are input to the D-Wave annealer to define
\begin{equation}
    f(\beta) = (\beta h_1 - \lambda_1)^2 + (\beta h_2 - \lambda_2)^2 + (\beta J -\lambda_2)^2,
    \label{eq:app_meth1_5}
\end{equation}
and minimize Eq.~(\ref{eq:app_meth1_5}) with respect to $\beta$, which yields
\begin{equation}
    \beta = \frac{\lambda_1^2+\lambda_2^2+\lambda_3^2}{h_1\lambda_1 + h_2\lambda_2 + J\lambda_3}.
\end{equation}
The estimated effective model parameters obtained from the D-Wave data are then given by $\widehat{h}_1= \lambda_1/\beta$, $\widehat{h}_2=\lambda_2/\beta$, and $\widehat{J}=\lambda_3/\beta$.

\textbf{Method 2:} Instead of using the spin averages and the correlations as the constraints, an alternative is to employ the average energy empirically obtained from the D-Wave data. Assuming the specified parameters to be the same as the ones implemented by the D-Wave annealers, the application of the principle of maximum entropy with $E_1(x_i) = h_1 S_1 + h_2 S_2 + J S_1 S_2$ and $K=1$ yields
\begin{align}
    &h_1 \widehat{S}_1 + h_2 \widehat{S}_2 + J \widehat{S}_{12}  =\nonumber \\
    &\frac{\sum_{S_1,S_2=\pm 1}(h_1 S_1 + h_2 S_2 + J S_1 S_2) e^{-\beta (h_1 S_1 + h_2 S_2 + J S_1 S_2)}}{\sum_{S_1,S_2=\pm 1}e^{-\beta (h_1 S_1 + h_2 S_2 + J S_1 S_2)}},
    \label{eq:app_method2}
\end{align}
where $\lambda_1 = \beta$ can be extracted by solving the non-linear Eq.~(\ref{eq:app_method2}).

We use these two methods to extract the parameters for three 2-variable instances. These are 2S4 with $h_1=-0.07$, $h_2=0.05$, and $J=0.1$ and previously-mentioned instances 2S1 and 2S2. 

In Table \ref{tab:piavgfirst}, we show the values of the extracted parameters by averaging over the 450 samples for each of the 2807 sequential (using only one pair of qubits at a time) runs  of the 2-variable problems on all 2807 available pairs of qubits on the D-Wave Advantage\_4.1. For 2S4, both Methods yield very similar values for $\beta$ and $T$.

For instances 2S1 and 2S2, it is noted that the theoretical equilibrium probability for sampling state $\ket{\downarrow\downarrow}$ at $\beta \approx 7$ is $\mathcal{O}(10^{-13})$. For these problems, the statistical error on $f_4$ is $\mathcal{O}(10^{-3})$, which is much too large to reliably extract the parameters as indicated by the disparity between the values of the $\beta$'s and $T$'s extracted using Method 1 and Method 2.

\begin{table}
    \centering
\caption{Parameters extracted using Methods 1 and 2 for three different 2-spin instances.}
    \begin{tabular}{|c|c|c|c|c|c|c|c|} \hline 
         Instance&  \multicolumn{5}{|c|}{Method 1}&  \multicolumn{2}{|c|}{Method 2}\\ \hline 
         &  $h_1$&  $h_2$& $J$ &  $\beta$ &  $T$(mK)&  $\beta$ &  $T$(mK)\\ \hline 
         2S4&  -0.069&  0.046&  0.103&  5.64&  36.5 &  5.59&  36.9\\ \hline 
         2S1&  -1.025&  -1.022&  0.899&  2.58&  80.0&  6.36&  32.4\\ \hline 
         2S2&  -1.001&  -0.996 &  1.002 &  2.75 &  74.9 &  5.00 &  41.2\\ \hline 
    \end{tabular}
    \label{tab:piavgfirst}
\end{table}

\section{Simultaneous execution of multiple copies of 2-variable problems}
\label{app:simultaneous}

In appendix~\ref{app:maxent}, we presented the D-Wave results for various 2-spin problems by collecting the data using all available qubit pairs sequentially, i.e., by employing only one pair at a time. A more resource-efficient alternative is to run many copies of the problem on the required number of qubit pairs simultaneously. 

\begin{figure*}
\begin{minipage}{0.33\textwidth}
         \centering
         \includegraphics[width=\textwidth]{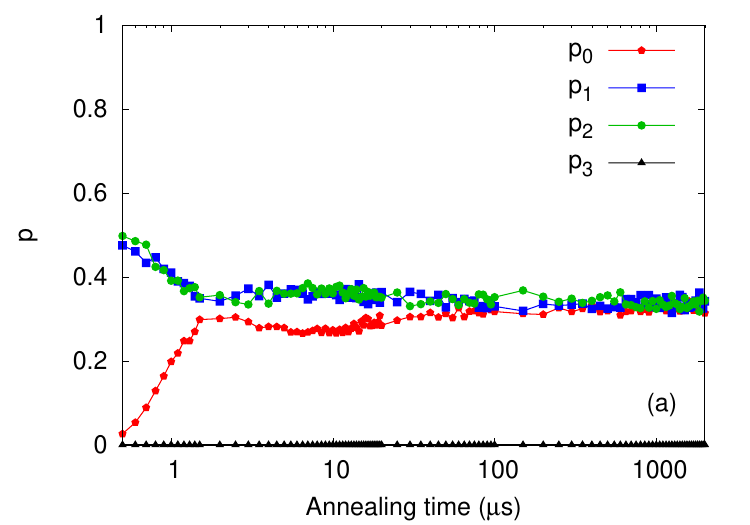}
         \put(-51, 102){\colorbox{white}{\makebox(12,2){\footnotesize $\ket{\uparrow\uparrow}$}}}
         \put(-51, 94){\colorbox{white}{\makebox(12,2){\footnotesize $\ket{\uparrow\downarrow}$}}}
         \put(-51, 86){\colorbox{white}{\makebox(12,2){\footnotesize $\ket{\downarrow\uparrow}$}}} 
         \put(-51, 78){\colorbox{white}{\makebox(12,2){\footnotesize $\ket{\downarrow\downarrow}$}}}
     \end{minipage}
     \hfill
     \begin{minipage}{0.33\textwidth}
         \centering
         \includegraphics[width=\textwidth]{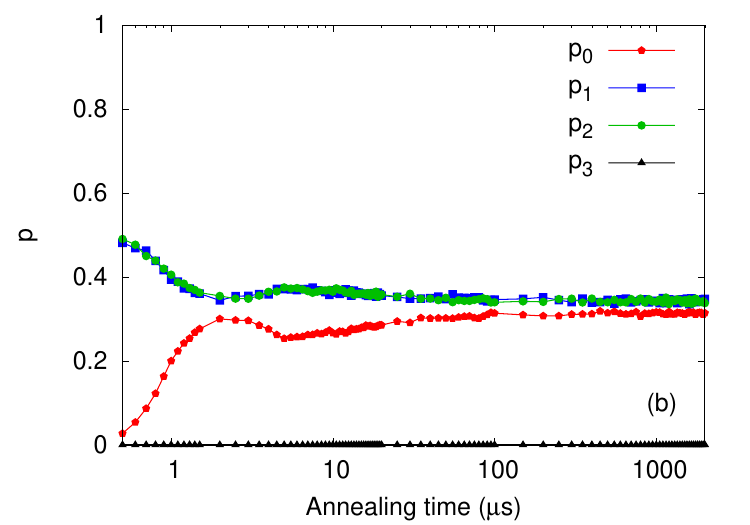}
         \put(-51, 102){\colorbox{white}{\makebox(12,2){\footnotesize $\ket{\uparrow\uparrow}$}}}
         \put(-51, 94){\colorbox{white}{\makebox(12,2){\footnotesize $\ket{\uparrow\downarrow}$}}}
         \put(-51, 86){\colorbox{white}{\makebox(12,2){\footnotesize $\ket{\downarrow\uparrow}$}}} 
         \put(-51, 78){\colorbox{white}{\makebox(12,2){\footnotesize $\ket{\downarrow\downarrow}$}}}
     \end{minipage} 
     \hfill
     \begin{minipage}{0.33\textwidth}
         \centering
         \includegraphics[width=\textwidth]{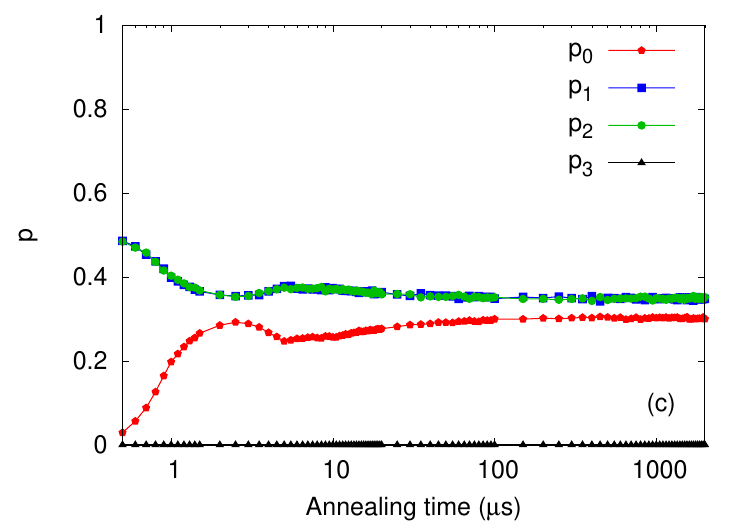}
         \put(-51, 102){\colorbox{white}{\makebox(12,2){\footnotesize $\ket{\uparrow\uparrow}$}}}
         \put(-51, 94){\colorbox{white}{\makebox(12,2){\footnotesize $\ket{\uparrow\downarrow}$}}}
         \put(-51, 86){\colorbox{white}{\makebox(12,2){\footnotesize $\ket{\downarrow\uparrow}$}}} 
         \put(-51, 78){\colorbox{white}{\makebox(12,2){\footnotesize $\ket{\downarrow\downarrow}$}}}
     \end{minipage} 
     \hfill
     \caption{(Color online) D-Wave data using the standard annealing protocol for 2-spin instance 2S2 running (a) 100, (b) 1000, and (c) 2500 copies of the problem simultaneously.}
    \label{fig:DW_J1_copies}
\end{figure*}
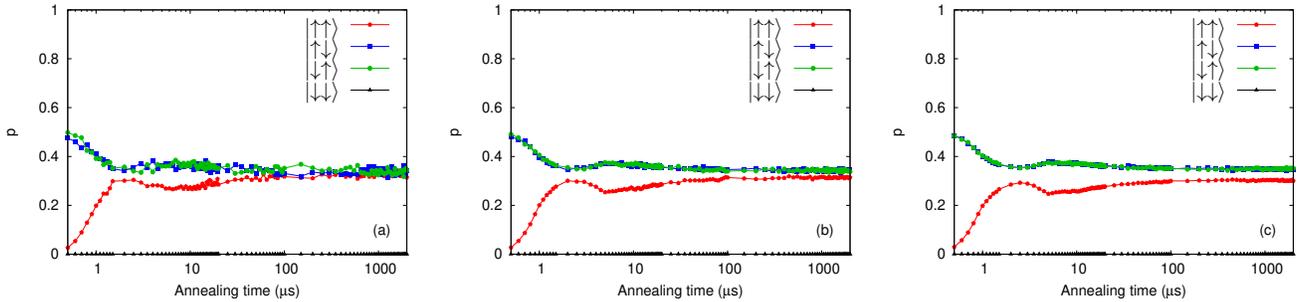

In Fig.~\ref{fig:DW_J1_copies}, we show the annealing time dependence of the probabilities for instance 2S2 for different numbers of copies. From the figure, it can be noted that as the number of copies, or in turn the number of qubits being simultaneously used, increases, the difference in the final probabilities for sampling state $\ket{\uparrow\uparrow}$ from those of states $\ket{\uparrow\downarrow}$ and $\ket{\downarrow\uparrow}$ increases systematically. The emergence of a reproducible, systematically increasing difference in the sampling probabilities can be regarded as an artefact, possibly arising from the bias created by the massive utilization of the qubits.

One way to remedy the above-mentioned artefact is to apply a spin-reversal transformation on the different copies, whereby we change the sign of the magnetic fields $h_i$'s for each spin with a certain probability and change the sign of the $J_{ij}$'s accordingly. 

\begin{figure}
    \centering
    \includegraphics[scale=0.6]{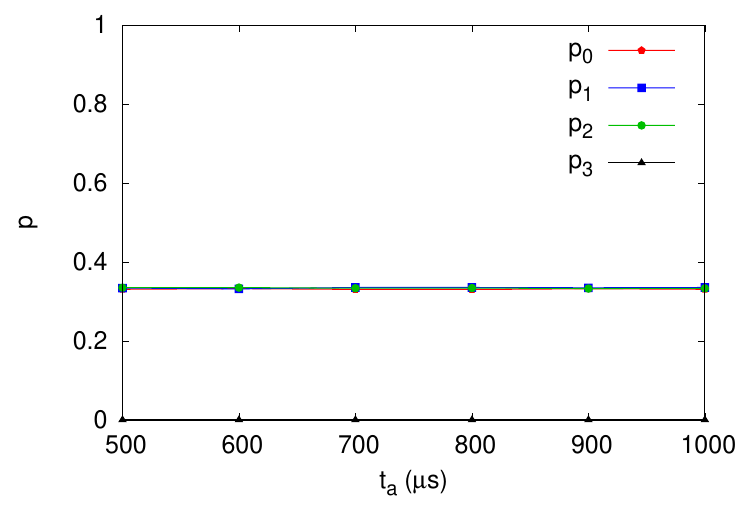}
    \put(-60, 135.3){\colorbox{white}{\makebox(12,2){\footnotesize $\ket{\uparrow\uparrow}$}}}
         \put(-60, 125){\colorbox{white}{\makebox(12,2){\footnotesize $\ket{\uparrow\downarrow}$}}}
         \put(-60, 114){\colorbox{white}{\makebox(12,2){\footnotesize $\ket{\downarrow\uparrow}$}}} 
         \put(-60, 103){\colorbox{white}{\makebox(12,2){\footnotesize $\ket{\downarrow\downarrow}$}}}
    \caption{(Color online) D-Wave data using the standard annealing protocol for 2-spin instance 2S2 running 2500 copies of the problem simultaneously with a spin reversal probability of 0.5.}
    \label{fig:DW_J1_spinrev}
\end{figure}

In Fig.~\ref{fig:DW_J1_spinrev}, we show the probabilities, averaged over 2500 copies of problem instance 2S1, with reversal probabilities $1/2$. From these results, it becomes clear that this treatment removes the artefact. This might suggest that the observed artefact in the annealing time scans originates from the net biases on the level of individual qubits in D-Wave systems and can, therefore, be eliminated by averaging the results for the copies after the spin-reversal transformation.

\section{Schr\"odinger dynamics of two spins interacting with a spin bath}
\label{app:spinbath}

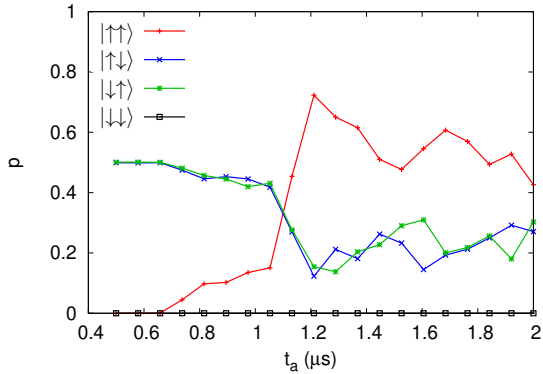
\begin{figure}
    \centering
    \includegraphics[scale=0.6]{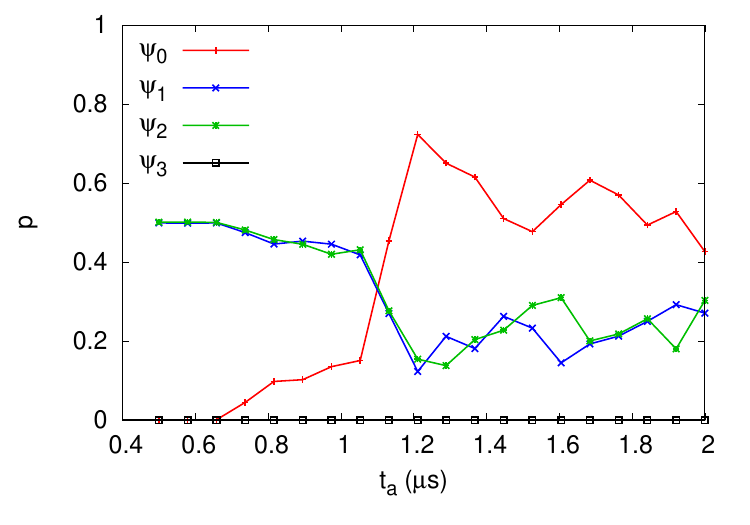}
    \put(-180.5, 135){\colorbox{white}{\makebox(14,2){\footnotesize $\ket{\uparrow\uparrow}$}}}
         \put(-180.5, 125){\colorbox{white}{\makebox(14,2){\footnotesize $\ket{\uparrow\downarrow}$}}}
         \put(-180.5, 114){\colorbox{white}{\makebox(14,2){\footnotesize $\ket{\downarrow\uparrow}$}}} 
         \put(-180.5, 103.5){\colorbox{white}{\makebox(14,2){\footnotesize $\ket{\downarrow\downarrow}$}}}
    \caption{(Color online) Data obtained using a 2-spin system interacting with a spin bath for problem instance 2S1. The number bath spins $N_{\mathrm{B}}=16$, $g=0.001$, $\Omega=0.1$, $t_{onset}^{start}$=0, and $t_{onset}^{end}$=900~ns.}
    \label{fig:spinbath}
\end{figure}
Focusing on the ``dips-and-bumps'' phenomenon (see section~\ref{sec:motivation}), in this appendix, we consider a microscopic model in which the two-qubit system interacts with a bath (B) of pseudo-spins (impurities, other qubits ...)
and question if the mechanism of the turn-on of the magnetic field proposed in the section~\ref{sec:sim_2spin} also explains the appearance of the  ``dips-and-bumps'' on the microscopic level.

The Hamiltonian of the system (S) + bath (B) takes the generic form
\begin{eqnarray}
\mathcal{H}(t)&=& H(t) + H_\mathrm{B} + g H_\mathrm{SB}
,
\label{SCB0}
\end{eqnarray}
where $H(t)$ is defined in Eq.~\ref{eq:DWHamil} and 
$H_\mathrm{B}$ and $H_\mathrm{SB}$ are the bath and the system-bath Hamiltonians, respectively. The overall strength of the system-bath interaction is controlled by the parameter $g$. The Hamiltonian for the system-bath interaction is chosen to be
\begin{eqnarray}
H_\mathrm{SB}&=& \sum_{m=1}^{2}\sum_{n=1}^{N_{\mathrm{B}}} \;\;\sum_{\alpha=x,y,z}K_{nm}^\alpha I_{n}^\alpha \sigma_m^\alpha
,
\label{SCB2}
\end{eqnarray}
where $N_{\mathrm{B}}$ is the number of spins in the bath, the $K^\alpha_{n}$'s are uniform random couplings in the range $[-K,+K]$ and $I_{n}^\alpha$ is the $\alpha$-th component of the bath spin $\mathbf{I}_n$.
As the overall system-bath interaction strength is controlled by $g$, without loss of generality, we may set $K=1\,\hbox{GHz}$.
Note that according to Eq.~(\ref{SCB2}), each of the two system spins interacts with each of the $N_{\mathrm{B}}$ bath spins. For the bath Hamiltonian, we take
\begin{eqnarray}
H_\mathrm{B}&=& \sum_{n=1}^{N_{\mathrm{B}}}\;\;\sum_{\alpha=x,y,z} \Omega_{n}^\alpha I_{n}^\alpha
,
\label{SCB3}
\end{eqnarray}
where the $\Omega_{n}^\alpha$'s are uniform random fields in the range $[-\Omega,\Omega]$ ($\Omega$ in units of GHz). The Hamiltonian Eq.~(\ref{SCB3}) describes a collection of spin-1/2 objects interacting with local fields.

The time evolution of a closed quantum system defined by a time-dependent Hamiltonian Eq.~(\ref{SCB0}) is governed by the time-dependent Schr\"odinger equation (TDSE)
\begin{eqnarray}
i\hbar\frac{\partial}{\partial t} |\Psi(t)\rangle &=& \mathcal{H}(t)|\Psi(t)\rangle
.
\label{QDWS0a}
\label{TDSE}
\end{eqnarray}%
The pure state $|\Psi(t)\rangle$ of the system-bath at time $t$ can be written as
\begin{eqnarray}
|\Psi(t)\rangle
&=&
\sum_{i=1}^4 \sum_{p=1}^{D_{\mathrm{B}}} c(i,p,t)|i,p\rangle
,
\label{QDWS0}
\end{eqnarray}%
where the complete set of the orthonormal states in up--down basis of the system and bath spins is denoted by $\{ |i,p\rangle \}$, $i$ labels the four states of the two system spins, and $D_{\mathrm{B}}=2^{{N_\mathrm{B}}}$ denotes the dimension of the Hilbert space of the bath. The coefficient $c(i,p,t)$ is the complex-valued amplitude of the state $|i,p\rangle$. From the knowledge of the $c$'s in Eq.~(\ref{QDWS0}), we can compute any physically relevant property of the system S, the bath B, and the whole system.

Given the random couplings, it is highly unlikely that  $H(t)$ and $H_\mathrm{SB}$ commute. Therefore, the system and the bath will exchange energy. Solving Eq.~(\ref{QDWS0a}) and tracing out the bath degrees of freedom implicitly accounts for non-Markovian processes and leads to decoherence and dissipation in a natural manner.

To account for the explicit time-dependence of $H(t)$, we use a second-order decomposition formula for ordered matrix exponentials~\cite{huyghebaert1990product,SUZU93,RAED06} to solve the TDSE Eq.~(\ref{QDWS0a}). The numerical method generates all $c(i,p,t)$'s in a time-stepping fashion. All TDSE simulation results reported in this section have been obtained by running a massively parallel, quantum spin dynamics simulator (in house software) with a time step of $10\,$ps for up to 200 000 time steps (depending on the annealing time $t_a$) and $N_\mathrm{B}=16$, the latter chosen to carry out the quantum annealing runs for different $t_a$, different choices of the parameters ($g$, $\Omega$, $h$-field onset parameters etc.) within a reasonable time span.

The initial state of the whole system is constructed using the random state technology~\cite{JIN21}. The key feature of random state technology is that if the dimension $D$ of Hilbert space is large enough, we can obtain an accurate estimate of $\mathbf{Tr}\;{\mathrm{X}}$, the trace of a matrix $X$, by computing $D\langle\Phi|X |\Phi\rangle$ where $|\Phi\rangle$ is a pure state chosen randomly~\cite{JIN21}. This reduces the calculation of averages by a factor $D$, which usually is a (very) large number.

In the case at hand, the initial state is taken to be a product state of the initial state of the system S and a pure random state of the bath B. The initial state of the system S is itself a product state of the two spins aligned along the $x$-direction (see section \ref{sec:methods}). The state of the whole system is then given by
\begin{eqnarray}
|\Psi(t=0)\rangle&=& |{}++\rangle \otimes |\Phi\rangle
\;,
\label{IS0}
\end{eqnarray}
where $|\Phi\rangle$ is a random state of the bath.

Figure~\ref{fig:spinbath} shows the time-dependent probabilities obtained from the simulation of the two spins interacting with the spin bath for problem instance 2S1. From these results, it is evident that even by following this general microscopic approach, it is possible to qualitatively capture the ``dips-and-bumps" feature observed in Fig.~\ref{fig:motiv_2spin}.

\section{Lindblad Master equation results with the standard annealing schedule}
\label{app:wo_onset}
As seen in section~\ref{sec:motivation}, the D-Wave results for the 2-spin problems show the so-called ``dips-and-bumps". In section~\ref{sec:sim_2spin}, we reproduce this feature by implementing different annealing functions for the linear and quadratic terms of the problem Hamiltonian in our simulations. Figure~\ref{fig:2spin_woonset} shows the result for the problem instance 2S1, without altering the annealing functions, from where it becomes evident that while these simulations capture the long annealing time behavior well, they do not produce the ``dips-and-bumps". 

\begin{figure}[H]
         \centering
         \includegraphics[scale=0.6]{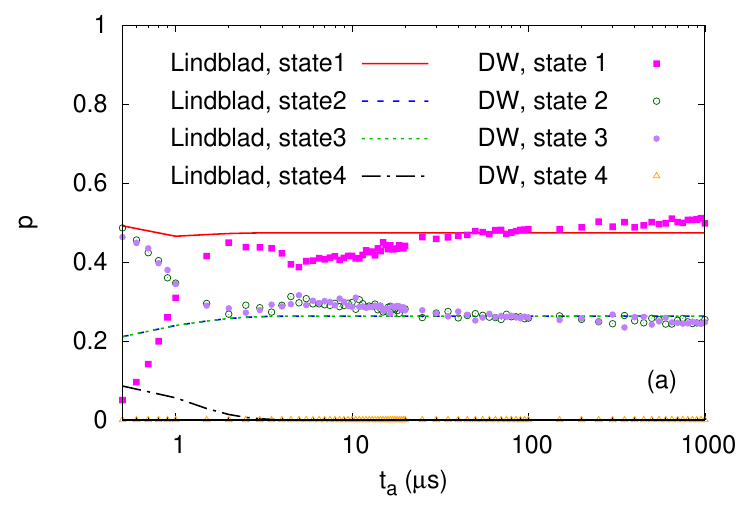}
        \put(-139, 132){\colorbox{white}{\makebox(18,2){\footnotesize $\ket{\uparrow\uparrow}$}}}
         \put(-139, 122){\colorbox{white}{\makebox(18,2){\footnotesize $\ket{\uparrow\downarrow}$}}}
         \put(-139, 111){\colorbox{white}{\makebox(18,2){\footnotesize $\ket{\downarrow\uparrow}$}}} 
         \put(-139, 100){\colorbox{white}{\makebox(18,2){\footnotesize $\ket{\downarrow\downarrow}$}}}
         \put(-65, 132){\colorbox{white}{\makebox(20,4){\footnotesize $\ket{\uparrow\uparrow}$}}}
         \put(-65, 122){\colorbox{white}{\makebox(20,2){\footnotesize $\ket{\uparrow\downarrow}$}}}
         \put(-65, 111){\colorbox{white}{\makebox(20,2){\footnotesize $\ket{\downarrow\uparrow}$}}} 
         \put(-65, 100){\colorbox{white}{\makebox(20,2){\footnotesize $\ket{\downarrow\downarrow}$}}}
     \caption{(Color online) Comparison of the D-Wave data with that from Lindblad master equation simulation with the standard annealing schedule for 2-spin instance 2S1 with $c=0.002$ and $T=35$~mK.}
    \label{fig:2spin_woonset}
\end{figure}
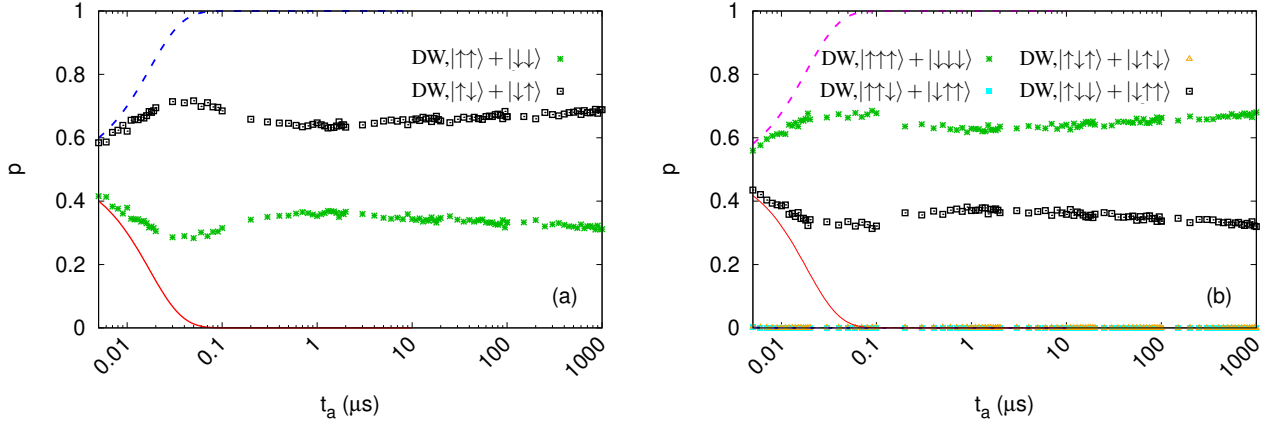
\begin{figure*}
     \begin{minipage}{0.49\textwidth}
         \centering
         \includegraphics[width=\textwidth]{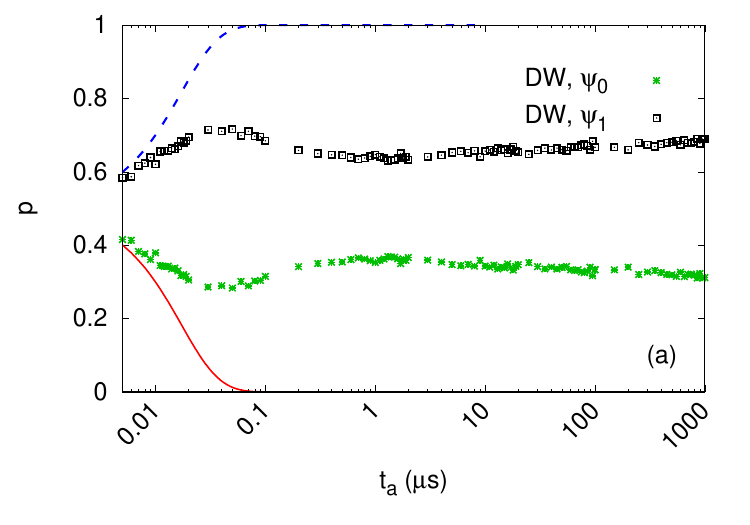}
         \put(-80, 144){\colorbox{white}{\makebox(30,3){\footnotesize DW,$\ket{\uparrow\uparrow}+\ket{\downarrow\downarrow}$}}}
         \put(-80, 131.5){\colorbox{white}{\makebox(30,3){\footnotesize DW,$\ket{\uparrow\downarrow}+\ket{\downarrow\uparrow}$}}}
     \end{minipage}
        \begin{minipage}{0.49\textwidth}
         \centering
         \includegraphics[width=\textwidth]{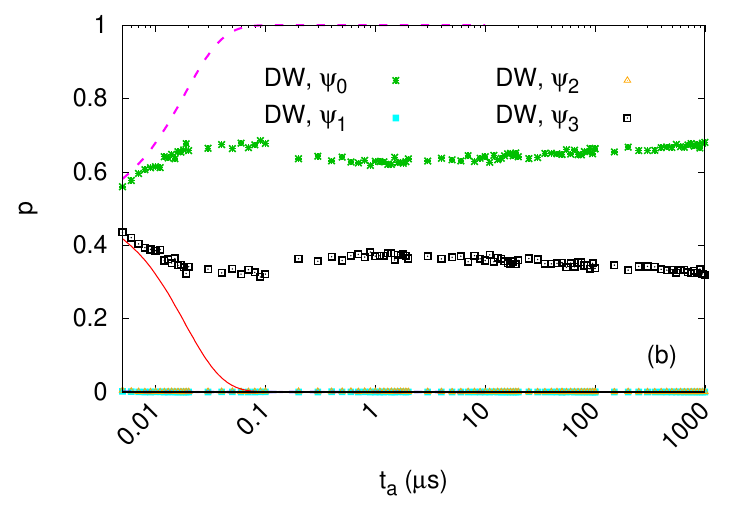}
         \put(-170, 143.5){\colorbox{white}{\makebox(33,3){\footnotesize DW,$\ket{\uparrow\uparrow\uparrow}+\ket{\downarrow\downarrow\downarrow}$}}}
         \put(-170, 131.5){\colorbox{white}{\makebox(32,3){\footnotesize DW,$\ket{\uparrow\uparrow\downarrow}+\ket{\downarrow\uparrow\uparrow}$}}}
         \put(-95, 143.5){\colorbox{white}{\makebox(33,3){\footnotesize DW,$\ket{\uparrow\downarrow\uparrow}+\ket{\downarrow\uparrow\downarrow}$}}}
         \put(-95, 131.5){\colorbox{white}{\makebox(33,3){\footnotesize DW,$\ket{\uparrow\downarrow\downarrow}+\ket{\downarrow\uparrow\uparrow}$}}}
     \end{minipage}
     \hfill
     \caption{(Color online) Comparison of D-Wave data using the fast annealing protocol with ideal quantum annealing simulations for (a) 2-spin problem instance with $J_{12}=0.05$ and (b) 3-spin problem instance with $J_{12}=1.00$, $J_{13}=-0.05$, and $J_{23}=-0.1$.}
    \label{fig:fast_sim_additional}
\end{figure*}
\section{Fast annealing results for additional problems}
\label{app:additional_fast}
Section~\ref{sec:fast_ann_sim} compares the D-Wave data for fast annealing with the corresponding simulations for problems involving non-zero $h_i$'s. We find the coherence times for these problems to be approximately 5~ns. In the present section, we do a similar comparison, but for problems with $h_i=0$. To this end, we choose a 2-variable problem with $J_{12}=0.05$ and a 3-variable problem with $J_{12}=1.00$, $J_{13}=-0.05$, and $J_{23}=-0.1$ such that no extra physical qubits are required. The results for these problems with the fast annealing feature do not suffer from any additional effects that might arise from using extra qubits to implement problems with non-zero $h_i$'s. The D-Wave data shown in Fig.~\ref{fig:fast_sim_additional} indeed shows lesser fluctuations compared to the data for the 2-spin instances with non-zero magnetic fields, shown in Fig.~\ref{fig:motiv_fast_2spin}. However, the D-Wave data shows quantitative agreement with the data from the \textcolor{blue}{TDSE} only up to approximately 5~ns.

\bibliography{references}
\end{document}